\documentclass[10pt,twocolumn]{IEEEtran}
\usepackage{amsmath}
\usepackage{amsfonts}
\usepackage{amssymb}
\usepackage{dsfont}
\usepackage{enumerate}
\usepackage{multirow}
\usepackage{color}
\usepackage{graphicx}
\usepackage{epstopdf}
\usepackage{caption}
\usepackage{cases}
\usepackage{subfigure}
\usepackage{cite}
\usepackage{bm}
\usepackage{algorithm}
\usepackage{algorithmicx}
\usepackage{algpseudocode}
\usepackage{setspace}
\usepackage{threeparttable}
\newtheorem{definition}{Definition}
\newtheorem{assumption}{Assumption}
\newtheorem{lemma}{Lemma}
\newtheorem{remark}{Remark}
\newtheorem{theorem}{Theorem}
\newtheorem{corollary}{Corollary}

\allowdisplaybreaks[4]

\begin{document}
\title{Intelligent Reflecting Surface Aided Pilot Contamination Attack and Its Countermeasure}
\author{Ke-Wen Huang, and Hui-Ming Wang, \emph{Senior Member,~IEEE}
\thanks{\scriptsize
The authors are with the School of Information and Communications
Engineering, and also with the Ministry of Education Key Laboratory for
Intelligent Networks and Network Security, Xi'an Jiaotong University, Xi'an
710049, China (email: {\tt xjtu-huangkw@outlook.com; xjbswhm@gmail.com}).
}
}

\IEEEtitleabstractindextext{
\begin{abstract}
Pilot contamination attack (PCA) in a time division duplex wireless communication system is considered, where an eavesdropper (Eve) attacks the reverse pilot transmission phase in order to wiretap the data transmitted from a transmitter, Alice, to a receiver, Bob. We propose a new PCA scheme for Eve, wherein Eve does not emit any signal by itself but uses an \emph{intelligent reflecting surface (IRS)} to reflect the pilot  sent by Bob to Alice.
The proposed new PCA scheme, referred to as IRS-PCA, increases the signal leakage from Alice to the IRS during the data transmission phase, which is then reflected by the IRS to Eve in order to improve the wiretapping capability of Eve.
The proposed IRS-PCA scheme disables many existing countermeasures on PCA due to the fact that with IRS-PCA, Eve no longer needs to know the pilot sequence of Bob, and therefore, poses severe threat to the security of the legitimate wireless communication system. In view  of this, the problems of 1) \emph{IRS-PCA detection} and 2) \emph{secure transmission under IRS-PCA} are considered in this paper.
For IRS-PCA detection, a generalized cumulative sum (GCUSUM) detection procedure is proposed based on the framework of \emph{quickest detection}, aiming at detecting the occurrence of IRS-PCA as soon as possible once it occurs.
For secure transmission under IRS-PCA, a cooperative channel estimation scheme is proposed to estimate the channel of the IRS, based on which zero-forcing beamforming is designed to reduce signal leakage.
\end{abstract}
}
\maketitle

\IEEEdisplaynontitleabstractindextext
\IEEEpeerreviewmaketitle

\section{Introduction}
Accurate channel state information (CSI) at transmitter side (CSIT) is a key enabler of physical layer secure transmission in multiple-antenna  wireless communication systems \cite{B.HeZTECom2013}.
In time division duplex (TDD) systems, CSIT is usually obtained by reverse pilot transmission (RPT) based on channel reciprocity.
However, the pilot sequence is usually publicly known, which gives rise to the so-called pilot contamination attack (PCA).

PCA was first proposed in \cite{X.ZhouTWC2012}, wherein an active eavesdropper (Eve) transmits the same pilot sequence as that transmitted by a legitimate receiver, Bob, to the legitimate transmitter, Alice, during the RPT phase. The occurrence of PCA causes inaccurate CSIT and leads to the significantly increased signal leakage to Eve during the subsequent data transmission (DT) phase. The vulnerability of multiple-antenna systems to PCA has been extensively investigated in \cite{Y.WuTIT2016,Y.O.BasciftciTIT2018,K.-W.HuangTWC2018,B.AkgunTIFS2019,D.KudathanthirigeTCOM2019}. Specifically, \cite{Y.O.BasciftciTIT2018} revealed that the secure degree of freedom of a single-cell cellular network becomes zero if an active Eve performs PCA. \cite{K.-W.HuangTWC2018} showed that if multiple Eves perform PCA cooperatively, the wiretapping signal-to-noise ratio (SNR) can be significantly improved.
\cite{Y.WuTIT2016,B.AkgunTIFS2019,D.KudathanthirigeTCOM2019} investigated PCA in multiple-cell and multiple-user networks, and the simulation results in \cite{B.AkgunTIFS2019} indicate that PCA degrades the network throughput by more than $50\%$.

\subsection{Related work and literature review}
In view of the severe threat of PCA on multiple-antenna systems, many methods have been proposed for PCA detection and secure transmission (ST) under PCA
\cite{Y.O.BasciftciTIT2018,Y.WuTIT2016,G.Zheng2013,X.WangVTC2017,
J.XieICC2017,W.ZhangTCOM2018,J.K.TugnaitWCL2017,J.K.TugnaitWCL2015,
J.K.TugnaitTCOM2017,W.WangTVT2019,X.TianAccess2017,D.HuWCL2019,X.HouWCSP2016,H.-MWang2018,
D.KapetanovicPIMRC2014,J.KangVTC2015,Y.WuTCOM2019,Q.XiongTIFS2016,Q.XiongTIFS2015,R.ZhuPIMRC2019,N.GaoSPL2019}.
Through a comprehensive review of existing literature, most existing countermeasures on PCA can be classified into one of the following four categories or be viewed as their combinations:
\emph{1) random modulation (RM) based method} \cite{G.Zheng2013,X.WangVTC2017,J.XieICC2017,W.ZhangTCOM2018},
\emph{2) artificial noise or random data (ANRD) aided method} \cite{J.K.TugnaitWCL2017,J.K.TugnaitWCL2015,J.K.TugnaitTCOM2017,W.WangTVT2019,X.TianAccess2017,D.HuWCL2019,Y.WuTCOM2019},
\emph{3) random orthogonal pilot (ROP) based method} \cite{X.HouWCSP2016,H.-MWang2018,Y.O.BasciftciTIT2018}, and
\emph{4) statistic feature (SF) based method} \cite{D.KapetanovicPIMRC2014,J.KangVTC2015,Y.WuTIT2016,Q.XiongTIFS2016,Q.XiongTIFS2015,R.ZhuPIMRC2019,N.GaoSPL2019}.
The common idea behind the methods in the former three categories is to introduce extra randomness into the standard RPT phase, which aims at disabling Eve to know in advance the sequence that Bob will send in the RPT phase.
The SF-based method relies on the fact that the SFs of the signal sequence received by Alice (Bob) in the RPT (DT) phase are quite different when PCA occurs and when PCA is absent.

Denote by $\bm{u}\in \mathcal{C}^{\tau_p\times 1}$ the publicly known pilot sequence with $\tau_p$ being the sequence length.
We briefly introduce the four categories of countermeasures on PCA in literature below.

\subsubsection{RM-based method \cite{G.Zheng2013,X.WangVTC2017,W.ZhangTCOM2018,J.XieICC2017}}
In RM-based method,  $\bm{u}$ is divided into several subsequences, e.g., $\bm{u} = [\bm{u}_1^T,\bm{u}_2^T,\cdots]^T$,  each of which is multiplied by a randomly generated symbol that is only known to Bob, and the sequence that Bob actually sends in the RPT phase is $\bm{u}_b(\bm{s}) = [s_1\bm{u}_1^T,s_2\bm{u}_2^T,\cdots]^T$, where $\bm{s} = [s_1,s_2,\cdots]^T$ consists of the random symbols.
To enable Alice to detect PCA, $\bm{s}$ is usually designed to present some special structure.
For instance, in \cite{G.Zheng2013}, $s_1,s_2,\cdots$ were randomly selected from a pre-given phase shift key constellation, and by utilizing the fact that $\bm{s}$ is unknown to Eve, PCA can be detected by checking the cross-correlation coefficient between two different channel estimations obtained by using two different pilot subsequences.
In general, if PCA occurs, the RM-based method does not enable Alice to estimate the CSI because Alice does not know $\bm{s}$. In \cite{J.XieICC2017},  $\bm{s}$ was secretly shared between Alice and Bob, and in this way, \cite{J.XieICC2017} proposed a method for Alice to simultaneously estimate the CSIs of Bob and Eve if PCA occurs.


\subsubsection{ANRD-aided method \cite{J.K.TugnaitWCL2017,J.K.TugnaitWCL2015,J.K.TugnaitTCOM2017,W.WangTVT2019,X.TianAccess2017,D.HuWCL2019,Y.WuTCOM2019}}
In ANRD-aided method, in addition to $\bm{u}$, Bob also sends a random sequence $\bm{s}\in\mathcal{C}^{\tau_s\times 1}$, which is only known to Bob.
There are two methods to transmit $\bm{s}$, which we refer to as \emph{superposition transmission} \cite{J.K.TugnaitWCL2017,J.K.TugnaitTCOM2017,J.K.TugnaitWCL2015} and \emph{separate transmission} \cite{W.WangTVT2019,D.HuWCL2019,X.TianAccess2017,Y.WuTCOM2019}, respectively.
In superposition transmission, $\bm{u}$ and $\bm{s}$ are simultaneously transmitted with $\tau_s=\tau_p$, and thus Bob actually sends $\bm{u}_b(\bm{s}) = \epsilon  \bm{u} + \sqrt{1 - \epsilon ^2} \bm{s}$ in the RPT phase,
where $\epsilon\in (0,1) $ is the power allocation factor.
In separate transmission, $\bm{s}$ is transmitted after the transmission of $\bm{u}$, and thus, the sequence that Bob actually sends is $\bm{u}_b(\bm{s}) = [\bm{u}^T, \bm{s}^T]^T  \in  \mathcal{C}^{(\tau_p+\tau_s)\times 1}$.
Due to the lack of knowledge about $\bm{s}$, the signal sequence sent by Eve will be linearly independent of $\bm{u}_b(\bm{s})$ with high probability. In view of this, \cite{J.K.TugnaitWCL2015} proposed a \emph{minimum description length} criterion based method to detect PCA. And if PCA truly occurs, \cite{J.K.TugnaitTCOM2017,D.HuWCL2019} proposed to estimate the channels of Bob and Eve by using \emph{independent component analysis}.


\subsubsection{ROP-based method \cite{X.HouWCSP2016,H.-MWang2018,Y.O.BasciftciTIT2018}}
In ROP-based method, the pilot transmitted by Bob is randomly selected from a set of mutually orthogonal pilots.
Eve not knowing which pilot Bob will send is not able to always transmit the same pilot as that transmitted by Bob.
Therefore, as long as Alice receives more than one pilot, the occurrence of PCA can be confirmed  \cite{X.HouWCSP2016}.
In \cite{Y.O.BasciftciTIT2018}, the authors proposed to encrypt the pilot so that Alice knows the pilot sequence sent by Bob in advance.
In this way, Alice can estimate the CSIs of Bob and Eve if PCA occurs.
In \cite{H.-MWang2018}, it was revealed that Alice and Bob need not encrypt the pilot as in \cite{Y.O.BasciftciTIT2018} if the channels of Bob and Eve were spatial correlated. In fact, knowing the channel covariance matrices, Alice can ``guess'' the pilot sent by Bob using maximal likelihood detection.

\subsubsection{SF-based method \cite{D.KapetanovicPIMRC2014,J.KangVTC2015,Y.WuTIT2016,Q.XiongTIFS2016,Q.XiongTIFS2015,R.ZhuPIMRC2019,N.GaoSPL2019}}
The occurrence of PCA leads to the increase in the energy received by Alice, denoted by $E_a$, in the RPT phase and the decrease in the energy received by Bob, denoted by $E_b$, in the DT phase, which motivates the energy-based detector in \cite{D.KapetanovicPIMRC2014,Q.XiongTIFS2015,R.ZhuPIMRC2019,N.GaoSPL2019}.
For instance, in \cite{D.KapetanovicPIMRC2014}, the occurrence of PCA was claimed if $E_a$ ($E_b$) was larger (small) than a predesigned threshold.
In \cite{Q.XiongTIFS2016}, except for the RPT phase, an extra forward channel training phase was used for Bob to estimate the CSI. If PCA does not exist, the CSI estimated by Bob is expected to be nearly the same as that estimated by Alice due to the channel reciprocity.
In \cite{J.KangVTC2015}, it was assumed that the channels of Bob and Eve were spatially correlated with $\bm{R}_b$ and $\bm{R}_e$ being the covariance matrices, respectively. Then, PCA can be detected by performing a (generalized) likelihood ratio test (LRT). In \cite{Y.WuTIT2016}, $\bm{R}_b$ and $\bm{R}_e$ were further utilized to construct minimum mean square error (MMSE) channel estimators and design secure beamforming vector.

\subsection{Motivations of this work}
Though the methods to combat with PCA have been extensively investigated for several years, the vulnerability of multiple-antenna systems to PCA is still far from being solved.

In this paper, we present a new PCA scheme for Eve which is shown to disable many existing countermeasures on PCA.
In the proposed PCA scheme, we let Eve exploit an \emph{intelligent reflecting surface}  (IRS) to perform PCA.
In brief, IRS is a passive device that does not emit any electromagnetic wave by itself.
Instead, it reflects the electromagnetic wave from the environment and is able to control the amplitudes and the phases of the reflecting coefficients.
Due to its flexibility, IRS has been widely considered as an approach to enhance the reliability and security of wireless communication systems by carefully selecting the reflecting coefficients, see e.g., \cite{C.Liaskos2018,Q.WuCMCM2020,LimengDongFx}.
While many researchers have proposed to utilize IRS to improve the performance of wireless communication systems, in this paper,
we reveal that IRS can also be used by Eve to deteriorate the performance of a wireless communication system.
In fact, with the aid of an IRS, Eve can attack the RPT phase of a TDD system effectively even without any knowledge on the pilot sequence.
Instead of transmitting some signal sequence by itself, Eve can simply deploy an IRS to reflect the pilot sequence transmitted by Bob to Alice.
As a result, during the RPT phase, the signal sequence from the IRS is always the same as that from Bob.
In principle, the RM-based, the ANRD-aided, and the ROP-based methods fail in combating with such an IRS-aided PCA (IRS-PCA) scheme because these methods require that the signal sequence sent by Eve is different from that sent by Bob.

Due to the significant differences between the IRS-PCA and the conventional PCA (C-PCA), i.e., Eve emits the pilot sequence by itself,
it is necessary to pay special attention to the IRS-PCA and investigate the corresponding countermeasures.
Basically, two issues are of great concern, i.e., 1) IRS-PCA detection, and 2) secure transmission under IRS-PCA.

\subsubsection{IRS-PCA detection}
Among the four categories of methods introduced in previous subsection, the SF-based method still facilitates the  detection.
However, the problem of PCA detection and the corresponding detection performance have not yet been fully illustrated in literature.
Specifically, in existing works, the problem of PCA detection was modeled as a binary hypothesis test (BHT) problem.
In each channel coherent time block, which consists of an RPT phase and a DT phase, either Alice or Bob makes a decision on whether PCA has occurred based on its received signals by performing a BHT. The detection process runs independently in different time blocks and the \emph{false alarm} and \emph{miss detection} probabilities in each single time block were used to characterize the detection performance.
However, in practice, if Eve launches PCA over multiple time blocks, then the detection performance can be significantly improved by combining the signals received in different time blocks together, which means that by performing the detection independently in different time blocks as in existing works may not be the optimal choice.


In practice, if Eve starts to attack the legitimate system at some moment, it is natural to require the legitimate system to quickly discover the presence of the attack so that countermeasures can be timely taken.
In fact, the optimal secure transmission schemes when PCA is absent and when PCA occurs are totally different.
When PCA is absent, the maximal ratio transmission (MRT) beamforming scheme maximizes the SNR of Bob.
But if PCA occurs, the MRT beamforming scheme causes significant signal leakage to Eve due to the inaccurate channel estimation.
To safeguard the data transmission under PCA, elaborated channel estimation and beamforming scheme should be used.
A basic question here is that how many time blocks are required for the legitimate system to detect the occurrence of PCA, i.e., the detection delay, subject to some constraint on the reliability of the detection result, so that the legitimate system can timely change its transmission scheme.
Besides, since PCA improves Eve's wiretapping capability, it is also relevant to know how many information bits that Eve can intercept by performing PCA before it being successfully detected by the legitimate system. These quantities are significant to characterize the efficiency of a PCA detection method, which, however, have not been studied in literature.

Motivated by the above observations, in this paper, we propose a new PCA detection method based on the theory of \emph{quickest detection} \cite{H.V.Poor2009,A.G.Tartakovsky2013}.
Our detection method works in a sequential manner, aiming at discovering the occurrence of IRS-PCA as quickly as possible once it starts.
At each time block, the signal sequence received by Alice during the RPT phase will be combined with all the signal sequences received in the past time blocks together to make a decision.
We analyze the detection delay of the proposed detection method and evaluate the extra amount of information that Eve can intercept due to its IRS-PCA from the time when IRS-PCA starts to the time when IRS-PCA is successfully detected.

\subsubsection{Channel estimation and secure transmission under IRS-PCA}
Under PCA, being able to estimate the channels of Bob and Eve is important for secure transmission.
In principle, the RM-based, the ANRD-aided, and the ROP-based methods do not work under the condition that Eve performs IRS-PCA.
For the SF-based method, some prior knowledge about the probability distribution of Bob's and Eve's channels is required to estimate the instantaneous channel coefficients. For example, in \cite{Y.WuTIT2016,H.-MWang2018}, the channel covariance matrices of Bob and Eve are required in order to construct linear MMSE channel estimators. However, in practice, it is not easy to obtain such channel statistics, especially that of Eve, because Eve should not cooperate with the legitimate system.

Based on the above observations, in this paper, we propose a new channel estimation scheme for Alice to estimate the channel between Alice and the IRS under the condition that Eve performs IRS-PCA. In the proposed channel estimation scheme, except for Bob, multiple cooperative nodes (CNs) also participate in the RPT process, each of which transmits a mutually orthogonal pilot sequence.
The basic idea is that the pilot sequences transmitted by the CNs will also be reflected by the IRS, and thus estimating the channel between Alice and the IRS can be achieved by analyzing the common component in the estimated channels of the multiple CNs.
The proposed scheme does not require Alice to know the covariance matrix of the channel of the IRS, which differs from the methods in \cite{Y.WuTIT2016,H.-MWang2018}.
With the proposed channel estimation scheme, zero-forcing (ZF) beamforming can be designed to null out the signal leakage, which greatly improves the secrecy performance of the legitimate system.


The rest of paper is organized as follows:  Section II introduces the system model and the proposed IRS-PCA; Section III discusses the proposed IRS-PCA detection scheme; Section IV presents the cooperative channel estimation and secure transmission scheme; Numerical results are presented in Section V; Finally, Section VI concludes the paper.

\emph{Notations:} $(\cdot)^*$, $(\cdot)^T$, and $(\cdot)^H$ denote conjugate, transpose, and conjugate transpose, respectively.
$\mathbb{E}(\cdot)$ and $\mathbb{P}(\cdot)$ denote mathematical expectation and probability, respectively.
$|\cdot|$ and $\|\cdot\|$ denote the absolute value and the norm, respectively.
$\mathcal{CN}(\cdot,\cdot)$ and $\mathcal{G}(\cdot,\cdot)$ denote the complex Gaussian and Gamma distributions, respectively.
$\mathrm{vec}(\bm{X})$ stacks the columns of $\bm{X}$ into a vector.
Diagonal matrix is denoted by $\mathrm{diag}(\cdot)$.
$\bm{I}_m$ denotes the $m$-by-$m$ identity matrix.
$\inf$, $\sup$, and $\mathrm{esssup}$ denote the infimum, the supremum, and the essential supremum, respectively.
$f(x) = \mathcal{O}(g(x))$ means that $\lim_{x\rightarrow\infty} \frac{f(x)}{g(x)}\leq c$ for some constant $c>0$. $f(x) = o_x(1)$ means that $\lim_{x\rightarrow \infty} f(x) = 0$.

\section{System Model and IRS-PCA}


Consider the transmission from a multiple-antenna transmitter, Alice, to a single-antenna receiver, Bob, which works in a TDD mode.
The whole communication process lasts for multiple time blocks and each time block consists of an RPT phase followed by a DT phase.
Denote $\bm{Y}_k\in \mathcal{C}^{M\times \tau_p}$ as the signal received by Alice during the the $k$-th RPT phase (i.e., the RPT phase in the $k$-th time block) with $M$ being the number of antennas and $\tau_p$ being the length of the pilot. In the following, we first introduce the proposed IRS-PCA scheme and then present some basic assumptions in this paper.

\subsection{IRS-PCA scheme}
\begin{figure}[t]
  \centering
  \includegraphics[width=3 in]{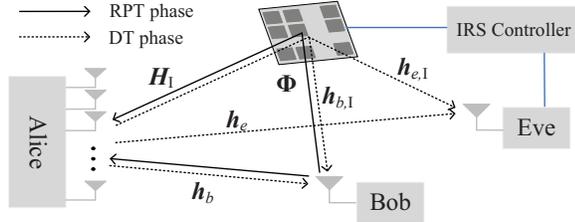}\\
  \caption{\small System Model.}\label{IRS-PCA-Model}
  \vspace{-5mm}
\end{figure}

In IRS-PCA, Eve keeps silent during the RPT phase. Instead, it utilizes an IRS, which consists of $N$ reflecting elements, to reflect the signal sequence sent by Bob to Alice.
During the $k$-th RPT phase,  Alice receives
\begin{align}
\bm{Y}_k &= \left\{
\begin{aligned}
&\underbrace{\sqrt{P_bg_b}  \bm{h}_{b,k} \bm{u}_{b,k}^H}_{\rm from~Bob} + \bm{Z}_k,  \  \text{if IRS-PCA does not occur},\\
&\underbrace{\sqrt{P_bg_b}  \bm{h}_{b,k} \bm{u}_{b,k}^H}_{\rm from~Bob} + \underbrace{\sqrt{P_bg_{\rm I}} \bm{H}_{{\rm I},k} \bm{\Phi}_k^{(p)} \bm{h}_{b,{\rm I},k} \bm{u}_{b,k}^H}_{\text{from the IRS}} \\
&\quad + \bm{Z}_k, \quad\quad\quad\quad\quad\ \  \text{if IRS-PCA  occurs},
\end{aligned}\right.
\label{IRSPCAAliceReceives}
\end{align}
where
$P_b$ is the transmit power of Bob,
$g_b$ and $g_{\rm,I}$ are the distance-based path losses from Bob and the IRS to Alice, respectively,
$\bm{h}_{b,k}\sim\mathcal{CN}(\bm{0},\bm{I}_M)$  and $\bm{H}_{{\rm I},k}$ denote the fading channels from Bob and the IRS to Alice in the $k$-th time block, respectively,
$\bm{h}_{b,{\rm I},k}$ denotes the channel from Bob to the IRS in $k$-th time block,
$\bm{\Phi}_k^{(p)} = \mathrm{diag} \{\phi_{1,k}^{(p)},\phi_{2,k}^{(p)}, \cdots,\phi_{N,k}^{(p)} \}\in\mathcal{C}^{N\times N}$ consists of the reflecting coefficients  of the IRS in the $k$-th RPT phase with $0\leq  |\phi_{i,k}^{(p)} |\leq 1$ for $1\leq i\leq N$,
$\left[\bm{Z}_k\right]_{i,j}\sim\mathcal{CN}(0,\sigma_a^2)$ for $1\leq i\leq M$ and $1\leq j\leq \tau_p$ denote the thermal noise received by Alice in the $k$-th RPT phase,
and $\bm{u}_{b,k}$ is the sequence that Bob sends in the $k$-th RPT phase.
\footnote{
In \eqref{IRSPCAAliceReceives}, the signals received by Alice are different when IRS-PCA occurs and when IRS-PCA is absent.
It should be noted that this does not necessarily mean that Eve needs to hide the IRS if the IRS-PCA is not harnessed.
In fact, in this paper, we do not impose any restriction on Eve's behavior when it does not perform the IRS-PCA.
}
We assume that all the channel coefficients are i.i.d. over different time blocks.
Note that $\bm{\Phi}_k^{(p)}$ is under the control of Eve.
In this paper, we assume that $\bm{\Phi}_k^{(p)}$ remains unchanged during the $k$-th RPT phase but can be different in different time blocks. Specifically, for $k^{\prime} \neq k$, $\bm{\Phi}_k^{(p)}$ is not necessarily equal to $\bm{\Phi}_{k^{\prime}}^{(p)}$. \footnote{
It is noteworthy that instead of randomly choosing the reflecting coefficients, Eve can carefully design the value of the reflecting coefficients in both the RPT and the DT phases in order to enhance its wiretapping capability. }
An comprehensive system model is illustrated in Fig \ref{IRS-PCA-Model}.

The effects of IRS-PCA can be summarized as follows:
\begin{enumerate}
\item It is of little use for the legitimate system to introduce any extra randomness to the RPT process because the signal sequence from Bob will always be reflected by the IRS, and Eve can effectively attack the RPT phase even without any knowledge about $\bm{u}_{b,k}$.
\item Alice cannot combat with IRS-PCA by utilizing the differences between the signal sequences from Bob and the IRS, and the RM-based, the ANRD-aided, and the ROP-based methods, in principle,  become disabled.
\end{enumerate}
Based on these observations, in this paper, we assume that Bob directly transmits the publicly known pilot sequence $\bm{u}$ in the RPT phase for simplicity, i.e., $\bm{u}_{b,k} = \bm{u}$.

\begin{figure*}
\small
\begin{minipage}[b]{\linewidth}
\centering
\begin{threeparttable}
\caption{\small COMPARISON BETWEEN DIFFERENT SOLUTIONS ON PCA.}\label{MethodComparision}
\begin{tabular}{|c|c|c|c|c|c|c|}
\hline\hline
Category
& \begin{tabular}[c]{@{}c@{}}Literature\end{tabular}
& \begin{tabular}[c]{@{}c@{}}Key Assumption\end{tabular}
& \begin{tabular}[c]{@{}c@{}}C-PCA\\detection\end{tabular}
& \begin{tabular}[c]{@{}c@{}}ST under\\C-PCA\end{tabular}
& \begin{tabular}[c]{@{}c@{}}IRS-PCA\\detection \end{tabular}
& \begin{tabular}[c]{@{}c@{}}ST under\\IRS-PCA \end{tabular} \\
\hline
\multirow{2}{*}{\begin{tabular}[c]{@{}c@{}} RM-based\\method\end{tabular}}
&\cite{G.Zheng2013,X.WangVTC2017,W.ZhangTCOM2018}& & $\surd$ & & &   \\
\cline{2-7}& \cite{J.XieICC2017}&\begin{tabular}[c]{@{}c@{}}$\bm{s}$ is secretly shared \\ between Alice and Bob \end{tabular}& $\surd$ & $\surd$  & &   \\
\hline
\multirow{4}{*}{\begin{tabular}[c]{@{}c@{}} ANRD-aided\\method\end{tabular}}
& \cite{J.K.TugnaitWCL2017,J.K.TugnaitTCOM2017,J.K.TugnaitWCL2015}& & $\surd$&$\surd$& & \\
\cline{2-7}& \cite{D.HuWCL2019,X.TianAccess2017} & \begin{tabular}[c]{@{}c@{}}$\bm{s}$ is (at least  partially) decodable\end{tabular}& $\surd$ & $\surd$  & $\surd$ & \\
\cline{2-7}& \cite{W.WangTVT2019} & \begin{tabular}[c]{@{}c@{}} $\bm{s}$ has a special structure\\that is unknown to Eve\end{tabular}& $\surd$ & $\surd$  & $\surd$ & \\
\cline{2-7}&\cite{Y.WuTCOM2019}& \begin{tabular}[c]{@{}c@{}}$\bm{R}_b$ is known to Alice, \\ Eve emits strong power\end{tabular}& -- & $\surd$ & $\surd$ & $\surd$ \\
\hline
\multirow{3}{*}{\begin{tabular}[c]{@{}c@{}} ROP-based\\method\end{tabular}}
           & \cite{X.HouWCSP2016}&\begin{tabular}[c]{@{}c@{}}Require more than one pilot\end{tabular}& $\surd$ & & & \\
\cline{2-7}& \cite{Y.O.BasciftciTIT2018}& \begin{tabular}[c]{@{}c@{}}$\bm{u}$ is secretly shared\end{tabular} & -- & $\surd$ & & \\
\cline{2-7}& \cite{H.-MWang2018}& \begin{tabular}[c]{@{}c@{}}Require more than one pilot,\\ $\bm{R}_b$ and $\bm{R}_e$ are known to Alice \end{tabular} & $\surd$ & $\surd$  & $\surd$ & $\surd$ \\
\hline
\multirow{4}{*}{\begin{tabular}[c]{@{}c@{}} SF-based\\method\end{tabular}}
           &\cite{D.KapetanovicPIMRC2014,R.ZhuPIMRC2019,N.GaoSPL2019}& & $\surd$ & &  $\surd$   & \\
\cline{2-7}&\cite{Q.XiongTIFS2015}&\begin{tabular}[c]{@{}c@{}} $E_a$ ($E_b$) is forwarded\\ (feedback) to Bob (Alice) \end{tabular}&$\surd$& $\surd$ & & \\
\cline{2-7}&\cite{Q.XiongTIFS2016}&\begin{tabular}[c]{@{}c@{}}Two-way training\end{tabular} & $\surd$ & $\surd$ & & \\
\cline{2-7}&\cite{J.KangVTC2015}&\begin{tabular}[c]{@{}c@{}}$\bm{h}_{\rm X}\sim\mathcal{CN}(\bm{0},\bm{R}_{\rm X})$,  for ${\rm X}\in\{b,e\}$ \end{tabular}  & $\surd$ & & $\surd$ & \\
\cline{2-7}&\cite{Y.WuTIT2016}& \begin{tabular}[c]{@{}c@{}}$\bm{R}_b$ and $\bm{R}_e$ are known to Alice\end{tabular}& -- & $\surd$ & $\surd$ & $\surd$ \\
\hline
\hline
\end{tabular}
\begin{tablenotes}
\item[1] {\footnotesize ``--'' means that the problem of C-PCA detection was not considered in the corresponding references. These works designed secure transmission schemes assuming that C-PCA always exists.}
\item[2] {\footnotesize $\bm{R}_b$ should be replaced with $\bm{R}_{\rm I}$ below ``Key Assumption'' when IRS-PCA is considered.}
\item
\end{tablenotes}
\end{threeparttable}
\end{minipage}
\vspace{-6mm}
\noindent\rule[0.25\baselineskip]{\textwidth}{0.5pt}
\end{figure*}

Based on \eqref{IRSPCAAliceReceives}, if the probability distribution of $\bm{h}_{b,k}$ is known, then (generalized) LRT facilitates to detect IRS-PCA, which degrades to an energy-based detector when $\bm{h}_{b,k}\sim\mathcal{CN}(\bm{0},\bm{I}_M)$.
Besides, based on the signal model in \eqref{IRSPCAAliceReceives}, to discriminate Bob's and the IRS's channels, it is necessary that Bob's and the IRS's channels follow different probability distributions.
For example, under the spatially correlated channel model, i.e., $\bm{h}_{b,k}\sim\mathcal{CN}(\bm{0},\bm{R}_b)$ and $\mathrm{vec}\left(\bm{H}_{{\rm I},k}\right)\sim\mathcal{CN}(\bm{0},\bm{R}_{\rm I})$, Alice can construct linear MMSE channel estimators based on $\bm{R}_b$ and $\bm{R}_{\rm I}$ as in \cite{Y.WuTIT2016,H.-MWang2018}.
A comparison between different solutions to PCA in literature is presented in Table \ref{MethodComparision}.

\begin{remark}
It is worth noting that the IRS-PCA scheme is related to the pilot replay attack (PRA) scheme studied in \cite{Y.ZengJSTSP2016,J.K.TugnaitVTC2017}, wherein Eve acts as an amplify-and-forward (AF) relay to receive and immediately retransmit the pilot sent by Bob.
The major difference between PRA and IRS-PCA is that in PRA, not only the pilot sent by Bob but also the residual self-interference (RSI) of Eve is retransmitted, as per \cite[Eqn. 3]{J.K.TugnaitVTC2017}.
Note that the RSI can be viewed as an independent signal component from Eve (independent of Bob's pilot $\bm{u}$), which is exploited in \cite{J.K.TugnaitVTC2017} to detect PRA.
However, in the proposed IRS-PCA scheme, we do not have such an RSI term in \eqref{IRSPCAAliceReceives}, and thus the method in \cite{J.K.TugnaitVTC2017} does not apply to our scenario. Besides, the method proposed in this paper can also be viewed as an approach to cope with PRA.
\end{remark}

\subsection{Assumptions and problem statement}
In this paper, we aim to design detection scheme for Alice to discover the occurrence of IRS-PCA and further establish secure transmission scheme to safeguard the data transmissions.
Before presenting any useful result, we first give some basic assumptions and definitions.

\begin{assumption}
\label{IRSEveChannelAssumption}
{\em
The channels from Bob and Eve to the IRS and from the IRS to Alice, i.e., $\bm{h}_{b,{\rm I},k}$, $\bm{h}_{e,{\rm I},k}$, and $\bm{H}_{{\rm I},k}$, can be decomposed as $\bm{h}_{b,{\rm I},k} = h_{b,{\rm I},k} \bm{\omega}_b$, $\bm{h}_{e,{\rm I},k} = h_{e,{\rm I},k} \bm{\omega}_e$, and $\bm{H}_{{\rm I},k} = \bm{h}_{{\rm I},k}\bm{\omega}_a^H$, respectively. Here, $\bm{\omega}_b,\bm{\omega}_e,\bm{\omega}_a\in \mathcal{C}^{N \times 1}$ are deterministic and Eve has perfect knowledge about $\bm{\omega}_b,\bm{\omega}_e,\bm{\omega}_a$. Besides, we assume that $\mathbb{E}\left(|h_{e,{\rm I},k}|^2\right) = \sigma_{e,{\rm I}}^2$, $\mathbb{E}\left(|h_{b,{\rm I},k}|^2\right) = \sigma_{b,{\rm I}}^2$, and $\bm{h}_{{\rm I},k}\sim\mathcal{CN}(\bm{0},\bm{I}_M)$.}
\hfill $\blacksquare$
\end{assumption}

The channel model in this assumption is in fact the Kronecker channel model in \cite{J.JoungTVT2017,H.ShinTIT2003,W.RheeISIT2006,S.ChatzinotasTWC2009}.  This assumption holds true if there are few scatterers around the IRS, and in this case $\bm{\omega}_b$, $\bm{\omega}_e$, and $\bm{\omega}_a$ can be viewed as the steering vectors of the array comprised of the multiple reflecting elements. It should be noted that this assumption enhances the attacking capability of Eve because with the knowledge about $\bm{\omega}_b$, $\bm{\omega}_e$, and $\bm{\omega}_a$, Eve can design passive beamforming for the IRS, which not only makes Alice's channel estimation more inaccurate but also enhances the wiretapping SNR in the DT phase.

Consider the communication between Alice and Bob over multiple channel coherence time blocks.
For convenience, we denote $\mathcal{T}_k$, $k\geq 1$, as the $k$-th time block which consists of an RPT phase with length $\tau_p$ and a DT phase with length $\tau_d$.
We assume that Alice knows the value of $a_b\triangleq \sqrt{P_bg_b} $, and
for $k\geq 1$, Alice can obtain the least square estimation of $\bm{h}_{b,k}$, denoted by
$\bm{y}_k = \frac{1}{a_b \tau_p}\bm{Y}_k \bm{u}$.
Assume that starting from some time block \emph{unknown} to Alice, denoted by $\mathcal{T}_{\nu}$ ($\nu \geq 1$), Eve performs IRS-PCA, aiming at intercepting the data transmitted by Alice in the DT phase, then $\bm{y}_k$ can be written as
\begin{subnumcases}{\label{BasicIRSPCAQDModel} \bm{y}_k = \frac{1}{a_b \tau_p}\bm{Y}_k \bm{u} = }
\bm{h}_{b,k} + \bm{z}_k,\quad\quad\quad\quad\quad k<\nu, &$~$, \label{BasicIRSPCAQDModelNoAttack}\\
\bm{h}_{b,k} + \hat{a}_{{\rm I},k} \bm{h}_{{\rm I},k} + \bm{z}_k,~~ k\geq\nu, &$~$, \label{BasicIRSPCAQDModelUnderAttack}
\end{subnumcases}
where $\hat{a}_{{\rm I},k} \triangleq \frac{a_{{\rm I},k}}{a_b}$ with $a_{{\rm I},k} \triangleq \sqrt{P_b g_{\rm I}} \bm{\omega}_a^H \bm{\Phi}_k^{(p)} \bm{\omega}_bh_{b,{\rm I},k}$ and
$\bm{z}_k = \frac{\bm{Z}_k\bm{u}}{a_b \tau_p}\sim \mathcal{CN}(\bm{0},\frac{\sigma_a^2}{a_b^2 \tau_p}\bm{I}_M)$.
Note that in \eqref{BasicIRSPCAQDModel}, if $\nu = 1$, it means that Eve starts to perform IRS-PCA at same time when Alice starts to communicate with Bob. Moreover, if $\nu = \infty$, it means that Eve never performs IRS-PCA.
For notational simplicity, in the following part of this paper, we use $\mathbb{P}_{k}\{\cdot\} = \mathbb{P} \{\cdot | \nu = k\}$ and $\mathbb{E}_{k}\{\cdot\} = \mathbb{E} \{\cdot | \nu = k\}$ to denote the probability and the mathematica expectation under the condition that $\nu = k$ for $\forall k\geq 1$.

In practice, if IRS-PCA does not exist, then the optimal beamforming scheme is the MRT scheme, which maximizes the SNR of Bob,  and the beamforming vector is given by
$\bm{w}_{\rm mrt} = \frac{\bm{y}_k}{\|\bm{y}_k\|}$.
However, if IRS-PCA occurs, using the MRT beamforming scheme leads to significant signal
leakage to the IRS because in this case, $\bm{y}_k$ is no longer an accurate estimation of Bob's channel.
The IRS can further reflect the signal from Alice to Eve, which greatly enhances the wiretapping SNR of Eve.
Therefore, using the MRT scheme under IRS-PCA will not be the best choice for securing the data transmission.
In view of the fact that Alice does not know when Eve starts to perform IRS-PCA, i.e., the value of $\nu$, we propose a sequential detection scheme for Alice to quickly determine whether IRS-PCA has occurred based on the framework of \emph{quickest detection} \cite{H.V.Poor2009}.
Theoretically, the proposed detection procedure can be written as a \emph{stopping time}, denoted by $T$, at which Alice declares the occurrence of IRS-PCA,
\begin{align}
\label{DetectionProcedure}
T = \inf\{k:k\geq 1,W_{k}(\bm{y}_1,\bm{y}_2,\cdots,\bm{y}_k) > \eta\},
\end{align}
where $W_{k}$ is the detection statistic in $\mathcal{T}_k$, which is a function of all past observations, and $\eta$ is the detection threshold.
Based on $T$ defined above, we make the following assumption on Alice's transmit beamforming scheme.
\begin{assumption}
\label{NaiveTransmissionScheme}
{\em Before the $T$-th time block, Alice views $\bm{y}_k$ as an estimation of $\bm{h}_{b,k}$ and uses $\bm{w}_{\rm mrt} = \bm{y}_k/\|\bm{y}_k\|$ as the beamforming vector to transmit its data to Bob.}
\end{assumption}

Note that Assumption \ref{NaiveTransmissionScheme} is natural because $\bm{w}_{\rm mrt}$ is optimal if no IRS-PCA exists and Alice does not have enough evidence to declare the occurrence of IRS-PCA before the $T$-th time block.
Under Assumption \ref{IRSEveChannelAssumption} and \ref{NaiveTransmissionScheme}, in the DT phase of $\mathcal{T}_k$, $\nu\leq k < T$, Bob and Eve receives
\begin{align}
\bm{y}_{b,k} &= \underbrace{\left( \sqrt{P_ag_b} \bm{h}_{b,k}^H \bm{w}_{\rm mrt}+ \sqrt{P_ag_{\rm I}} \bm{q}_b^H\bm{w}_{\rm mrt} \right)}_{\triangleq G_{b,k}}\bm{x}_a^H + \bm{z}_{b,k}^H \\
\bm{y}_{e,k} &=  \left( \sqrt{P_ag_e} \bm{h}_{e,k}^H \bm{w}_{\rm mrt}+ \sqrt{P_ag_{\rm I}}  \bm{q}_{e,{\rm I}, k}^H  \bm{w}_{\rm mrt} \right)\bm{x}_a^H + \bm{z}_{e,k}\nonumber \\
&= \underbrace{\left( a_e \bm{h}_{e,k}^H + \tilde{a}_{{\rm I},k} \bm{h}_{{\rm I},k}^H\right)\bm{w}_{\rm mrt}}_{\triangleq G_{e,k}}\bm{x}_a^H + \bm{z}_{e,k}^H,
\label{SignalReceivedbyEve}
\end{align}
where $\bm{q}_{b,k}^H = \bm{h}_{b,{\rm I},k}^H\bm{\Phi}_k^{(d)} \bm{H}_{{\rm I},k}^H$,
$\bm{q}_{e,{\rm I}, k}^H = \bm{h}_{e,{\rm I},k}^H\bm{\Phi}_k^{(d)} \bm{H}_{{\rm I},k}^H$
$a_e\triangleq \sqrt{P_a g_e}$ with $P_a$ being the transmit power of Alice and $g_e$ being the path loss between Alice and Eve,
$\tilde{a}_{{\rm I},k}\triangleq \sqrt{P_ag_{\rm I}} h_{e,{\rm I},k}\bm{\omega}_{e}^H\bm{\Phi}_k^{(d)}\bm{\omega}_{a}$,
$\bm{x}_a\in\mathcal{C}^{\tau_d\times 1}$ denotes the data sequence that Alice transmits to Bob satisfying $\frac{1}{\tau_d}\mathbb{E}\left\{\|\bm{x}_a\|^2\right\}=1$,
$\bm{z}_{b,k}\sim \mathcal{C}(\bm{0},\sigma_b^2\bm{I}_{\tau_d})$ and $\bm{z}_{e,k}\sim \mathcal{C}(\bm{0},\sigma_e^2\bm{I}_{\tau_d})$ denote the noise received by Bob and Eve, respectively,
$\bm{h}_{e,k} \sim \mathcal{CN}(\bm{0},\bm{I}_M)$ is the channel from Eve to Alice,
and
$\bm{\Phi}_k^{(d)} = \mathrm{diag}\left\{\phi_{1,k}^{(d)},\phi_{2,k}^{(d)}, \cdots,\phi_{N,k}^{(d)}\right\}\in\mathcal{C}^{N\times N}$ consists of the reflecting coefficients of the IRS in the $k$-th DT phase.
Note that the same as $\bm{\Phi}_k^{(p)}$, $\bm{\Phi}_k^{(d)}$ is under the control of Eve.
We assume that $\bm{\Phi}_k^{(d)}$ remains unchanged in the $k$-th DT phase, but for $k^{\prime} \neq k$, $\bm{\Phi}_k^{(d)}$ can be different from $\bm{\Phi}_{k^{\prime}}^{(d)}$.
For simplicity, we assume that Bob and Eve know the value of $G_{b,k}$ and $G_{e,k}$, respectively. This is because both $G_{b,k}$ and $G_{e,k}$ are scalars, and thus can be easily learned by a dedicated training process or directly estimated from the received data sequence. Based on \eqref{SignalReceivedbyEve}, the amount of information that Eve can intercept in the $k$-th DT phase, $\nu\leq k < T$, is
\begin{align}
\mathcal{I}_k \triangleq \tau_d \times C_{e,k}^{\rm \text{IRS-PCA}} = \tau_d \times \ln\left(1 + \frac{|G_{e,k}|^2}{\sigma_e^2}\right),
\label{AWTBasic}
\end{align}
where $C_{e,k}$ is the channel capacity of Eve based on the signal model in \eqref{SignalReceivedbyEve}.
Based on \eqref{AWTBasic}, we define the \emph{wiretapping throughput gain (WTG)} below, which characterizes the extra amount of information that Eve can eavesdrop due to its IRS-PCA.
\begin{definition}
\label{AWTGDefinition}
{\em Define ${\rm WTG}_n \triangleq \sum_{k=\nu}^{n} \left( \mathcal{I}_k - \tilde{\mathcal{I}}_k \right)
= \sum_{k=\nu}^{n}  \Delta\mathcal{I}_k$, where
$\tilde{\mathcal{I}}_k \triangleq \tau_d \times C_{e,k}^{\rm no~IRS}
=  \tau_d \times \ln\left(1 + \frac{|a_e|^2}{\sigma_e^2}\frac{ \left|\bm{h}_{e,k}^H\left(\bm{h}_{b,k} + \bm{z}_k\right)\right|^2}{\left\|\bm{h}_{b,k} + \bm{z}_k\right\|^2}\right)$
is the amount of information that Eve can intercept during $\mathcal{T}_k$ for $k\geq \nu$ under the condition that no IRS exists and Eve does not launch any active attack to the legitimate system.
}
\end{definition}

\begin{figure*}[t]
  \centering
  \includegraphics[width=4.5 in]{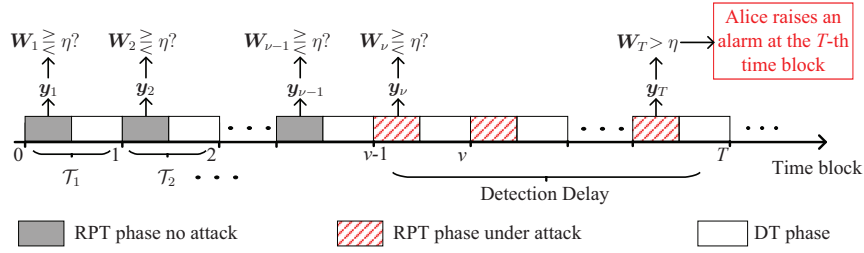}\\
  \caption{\small Detection Model.}\label{Detection-Model}
  \vspace{-4mm}
\end{figure*}


In general, a single run of the detection process in \eqref{DetectionProcedure} possibly generates two results.
A \emph{false alarm event} occurs if $T<\nu$. Recall that it is possible that Eve does not perform IRS-PCA, i.e., $\nu=\infty$, and in this case, any alarm raised by Alice's detector is false. \emph{Successful detection} occurs if $T\geq\nu$, and in this case, $T-\nu$ is referred to as the detection delay. We illustrate an example of the detection process in Fig. \ref{Detection-Model}.

In this paper, we are interested in the following three quantities which can be used to characterize the performance of the detection procedure in the form of \eqref{DetectionProcedure}.
\begin{enumerate}
\item \emph{Average run length to false alarm (ARL2FA):} Under the condition that Eve never performs IRS-PCA, the mean value of $T$ is referred to as   ARL2FA, i.e., ${\rm ARL2FA} = \mathbb{E}_{\infty}(T)$.
\item \emph{Worst-case average detection delay (WADD):} Denoted by $\mathcal{D}(T)$, WADD is defined as
    \begin{align}
    \label{WADD}
    \mathcal{D}(T) &\triangleq \sup_{\nu\geq1} \mathcal{D}_\nu(T)
    \end{align}
    where $\mathcal{D}_\nu(T) \triangleq \mathop{\mathrm{esssup}}_{\bm{y}_1^{\nu-1}}~\mathbb{E}_{\nu}  ( (T-\nu)^+ | \bm{y}_1^{\nu-1})$ and $\bm{y}_1^{\nu-1} = (\bm{y}_1,\bm{y}_2,\cdots,\bm{y}_{\nu-1})$.
\item \emph{Worst-case average wiretapping throughput gain (WAWTG):} compared with the case where there is no IRS in the system, ${\rm WTG}_n$, defined in Definition \ref{AWTGDefinition}, can be viewed as the extra amount of data that Eve can intercept by performing IRS-PCA from the $\nu$-th to the $n$-th time block. Accordingly, we define the WAWTG, $\mathcal{W}(T)$, as
    \begin{align}
    \mathcal{W}(T) &\triangleq \sup_{\nu\geq 1}\mathop{\mathrm{esssup}}_{\bm{y}_1^{\nu-1}}~\mathbb{E}_{\nu}( {\rm WTG}_{T-1} | \bm{y}_1^{\nu-1} ) \nonumber \\
    &= \sup_{\nu\geq 1}\mathop{\mathrm{esssup}}_{\bm{y}_1^{\nu-1}}~\mathbb{E}_{\nu}\left( \sum_{k=\nu}^{T-1}  \Delta\mathcal{I}_k \bigg| \bm{y}_1^{\nu-1} \right).
    \end{align}
    where $\sum_{k=\nu}^{\nu-1}(\cdot)$ is understood as zero.
\end{enumerate}

In practice, if IRS-PCA does not occur, then it is expected that Alice's detector raises few false alarms as time goes by, which requires the ARL2FA to be large.
If Eve indeed starts to perform IRS-PCA for some $\nu$ unknown to Alice, it is expected that Alice can quickly discover the occurrence of IRS-PCA so that Alice can timely take countermeasures, i.e., WADD is desired to be small.
In fact, the ARL2FA and the WADD are two important metrics to evaluate the performance a quickest detection scheme, see e.g. \cite[Section 6]{A.G.Tartakovsky2013}, and a widely used method to design quickest detection scheme is to minimize the WADD subject to a lower bound on the ARL2FA, for example, the \emph{cumulative sum} (CUSUM) detection procedure \cite{G.Lorden1971}.
In this paper, in addition to the ARL2FA and the WADD, we are also interested in evaluating the value of $\mathcal{W}(T)$.
In fact, $\mathcal{W}(T)$ can be viewed as a weighted version of the WADD, and the weight coefficients are $\Delta\mathcal{I}_k$ for $k\geq \nu$.
Note that $\mathcal{W}(T)$ can be viewed not only as a performance indicator of Alice's IRS-PCA detection scheme but also as a measure on the threat posed by Eve's attack.
On one hand, for a given attacking strategy of Eve, if Alice's detector can discover the existence of Eve's attack more quickly,
then the number of terms in the summation formula in Definition \ref{AWTGDefinition} becomes fewer, and thus $\mathcal{W}(T)$ will be smaller.
On the other hand, for a given IRS-PCA detection scheme of Alice, $\mathcal{W}(T)$ can be viewed as the secrecy performance loss of legitimate communication system due to Eve's attack, and a smaller value of $\mathcal{W}(T)$ means that Eve's attack causes less harm to the legitimate communication system.

Based on \eqref{BasicIRSPCAQDModel}, the detection performance is highly related to the sequence
$\{\hat{a}_{{\rm I},k}:k\geq \nu\}$. Though $\{\hat{a}_{{\rm I},k}:k\geq \nu\}$ is under the control of Eve, we make the following assumption.
\begin{assumption}
\label{AverageAttackPowerAssumption}
{\em Suppose
\begin{align}
\frac{1}{n\sigma_0^2}\sum_{j=k}^{k+n-1} |\hat{a}_{{\rm I},\nu + j}|^2
= \frac{1}{n }\sum_{j=k}^{k+n-1} \mu_{\nu+j} \xrightarrow{n\rightarrow\infty} \mu,
\label{AverageAttackPower}
\end{align}
uniformly in $k\geq 0$ for some $\mu\in (0,\infty)$, where $\sigma_0^2 \triangleq 1 + \frac{\sigma_a^2}{a_b^2 \tau_p}$ and $\mu_{\nu+j} \triangleq \frac{|\hat{a}_{{\rm I},\nu + j}|^2 }{\sigma_0^2}$.
}
\end{assumption}

Note that the IRS-PCA detection scheme proposed in this paper does not require Alice to know the value of $\mu$ and $\{\hat{a}_{{\rm I},k} : k \geq \nu\}$. But this assumption is essential for analyzing the WADD and the WAWTG. Besides, this assumption naturally holds if the reflecting coefficients of the IRS remain the same over different time blocks, i.e., $\bm{\Phi}^{(p)} = \bm{\Phi}_\nu^{(p)} = \bm{\Phi}_{\nu+1}^{(p)} = \cdots$, and in this case, $\mu = \frac{P_b g_{\rm I} |\bm{\omega}_a^H \bm{\Phi}^{(p)} \bm{\omega}_b |^2  }{a_b^2\sigma_0^2}\sigma_{b,{\rm I}}^2$.

Based on the above assumptions, in next section, we present an IRS-PCA detection scheme for Alice to determine whether IRS-PCA has occurred. Given that Eve performs IRS-PCA, a channel estimation and secure transmission scheme is presented in Section \ref{Sec:SecureTransmission} to safeguard the data transmissions from  Alice to Bob.

\begin{remark}
In formula \eqref{BasicIRSPCAQDModel}, it is implicitly assumed that once Eve starts attacking in $\mathcal{T}_{\nu}$, it continues to attack in the all subsequent time blocks, $\{\mathcal{T}_{\nu+l}:~l\geq 1\}$. It is straight to extend the model in \eqref{BasicIRSPCAQDModel} to more general case where whether IRS-PCA occurs or not follows some prior probabilities that are selected by Eve. For example, Eve can choose to perform IRS-PCA in $\mathcal{T}_{\nu+l}$ for $l\geq 1$ with probability $0<q_{l}<1$. And if for some $l\geq 1$, Eve chooses not to attack in $\mathcal{T}_{\nu+l}$, then $\hat{a}_{{\rm I},\nu+l}$ should be viewed as zero in \eqref{BasicIRSPCAQDModel} and \eqref{AverageAttackPower}.
Even so, the IRS-PCA detection scheme proposed in this paper still works if \eqref{AverageAttackPower} holds for some $\mu>0$.
\end{remark}
\begin{remark}
In practice, it is possible that Eve utilizes an IRS or an AF relay to only attack the DT phase. Specifically, the IRS (the AF relay) is only used to reflect (retransmit) the signal sent by Alice in the DT phase to enhance Eve's wiretapping SNR without attacking the RPT phase. In this situation, the channel estimation process of the legitimate system is not affected by Eve's attack, and therefore, the detection scheme presented in this paper does not facilitate the legitimate system to discover such kind of attack. Note that, in this paper, we focus on the situation where Eve uses the proposed  IRS-PCA scheme to contaminate the channel coefficients estimated by Alice. Detecting the attack mentioned above, which occurs in the DT phase only, is beyond the scope of this paper and constitutes an interesting future research.
\end{remark}

\section{Discovering IRS-PCA: A quickest detection framework}
\label{Sec:Detection}
In this section, we present our IRS-PCA detection method and analyze its performance.
According to \eqref{BasicIRSPCAQDModel}, if IRS-PCA never occurs,  then $\{\bm{y}_k: k\geq 1\}$ is a sequence of i.i.d. random variables with common distribution $\mathcal{CN}(\bm{0},\sigma_0^2\bm{I}_M)$ and $\sigma_0^2 = 1 + \frac{\sigma_a^2}{a_b^2 \tau_p}$. If there exists a finite $\nu$ such that Eve starts to perform IRS-PCA, then for $k \geq \nu$, the distribution of $\bm{y}_k$ becomes $\mathcal{CN}(\bm{0},(\sigma_0^2 + |\hat{a}_{{\rm I},k}|^2) \bm{I}_M)$.
In other words, if IRS-PCA occurs for some finite $\nu$, the sequence $\{\bm{y}_k : k\geq 1\}$ undergoes a change in its statistic distribution. In practice, it is essential for the legitimate system to discover the existence of the attack as soon as possible.
In view of this, the detection of IRS-PCA naturally falls into the field of  \emph{quickest detection} \cite{H.V.Poor2009}.

A widely used quickest detection scheme is the CUSUM procedure defined as follow,
\begin{align}
T_{\rm CU} = \inf \left\{ k: k\geq 1, \max_{1\leq t \leq k}\sum_{j=t+1}^k \Lambda(\bm{y}_j) > \eta_{\rm CU} \right\},
\end{align}
where $\Lambda(\bm{y}_k)$, $k\geq 1$, is the log-likelihood ratio and $\eta_{\rm CU}$ is the detection threshold.
In fact, if the post change samples, i.e., $\{\bm{y}_k : k\geq \nu\}$,  are i.i.d. distributed and if $\eta_{\rm CU}$ is selected such that $E_{\infty}(T_{\rm CU}) = \gamma > 1$, then the CUSUM procedure is optimal in minimizing the WADD with the ARL2FA no smaller than $\gamma$. Namely, the CUSUM procedure is optimal in the following sense \cite{G.Lorden1971,G.V.Moustakides1986},
\begin{align}
\label{QDOptimizationProblem}
\inf_{T} \mathcal{D}(T), \quad {\rm ~s.t.~} \mathbb{E}_{\infty}(T) \geq \gamma,
\end{align}
where $\mathcal{D}(T)$ is defined in \eqref{WADD}.

In our case, $f_{1,k}(\bm{y}_k) = \frac{\exp\left\{ - \|\bm{y}_k\|^2/(\sigma_0^2 + |\hat{a}_{{\rm I},k}|^2)\right\}}{\pi^M (\sigma_0^2 + |\hat{a}_{{\rm I},k}|^2)^M}$, $f_0(\bm{y}_k ) = \frac{\exp\left\{ - \|\bm{y}_k\|^2/\sigma_0^2 \right\}}{\pi^M \sigma_0^{2M} }  $, and thus
the CUSUM procedure becomes
\begin{align}
T_{\rm CU} &= \inf \Big\{ k: k\geq 1, \nonumber \\
&\quad \max_{1\leq t \leq k}\sum_{j=t+1}^k \left(A_k \|\bm{y}_k\|^2 - B_k M \right) > \eta_{\rm CU} \Big\}.
\end{align}
where $A_k \triangleq \frac{|\hat{a}_{{\rm I},k}|^2}{\sigma_0^2\left(\sigma_0^2 + |\hat{a}_{{\rm I},k}|^2\right)  }$ and $B_k \triangleq \ln \left(1 + \frac{|\hat{a}_{{\rm I},k}|^2}{\sigma_0^2}\right)$.
However, the CUSUM procedure cannot be directly applied in our case because $A_k$ and $B_k$ contain unknown parameters $\{ \hat{a}_{e,k} : k\geq \nu \}$. In the following, we propose a generalized CUSUM (GCUSUM) procedure to make it capable of detecting IRS-PCA.

The proposed GCUSUM procedure can be written as the following stopping time,
\begin{align}
\label{ProposedGCUSUM}
T_{\rm G} &= \inf \left\{ n: n\geq 1, \max_{1\leq k \leq n} \tilde{\Lambda}_{k,n} > \eta_{\rm G} \right\},\\
\tilde{\Lambda}_{k,n} &=
\sup_{x\geq \xi }\Bigg\{ \frac{x S_{k,n}}{\sigma_0^2 + x} -  M(n-k+1)\ln\left(1 + \frac{x}{\sigma_0^2}\right) \Bigg\} \nonumber\\
&=\left\{\begin{aligned}
& M(n-k+1) \left(\bar{S}_{k,n} - \ln \bar{S}_{k,n} - 1\right),  \\
&\quad\quad\quad\quad\quad\quad\quad\quad\quad\quad\quad\quad
\text{ if } \bar{S}_{k,n} - 1 \geq \bar{\xi}, \\
& \frac{\bar{\xi} M(n-k+1)\bar{S}_{k,n}}{ 1 + \bar{\xi} } \\
&-  M(n-k+1)\ln \left(1 + \bar{\xi}\right),   \text{ if } \bar{S}_{k,n} - 1 < \bar{\xi},
\end{aligned}\right.\nonumber
\end{align}
where $\eta_{\rm G}$ is the detection threshold, $\xi>0$ is a parameter to be designed,
$\bar{\xi} \triangleq \frac{\xi}{\sigma_0^2}$, $S_{k,n} \triangleq \frac{\sum_{j=k}^n \|\bm{y}_j\|^2}{\sigma_0^2} $, and $\bar{S}_{k,n} \triangleq \frac{\sum_{j=k}^n \|\bm{y}_j\|^2}{M(n-k+1)\sigma_0^2}$. The detection performance of \eqref{ProposedGCUSUM} is characterized in the following theorem.
\begin{theorem}
\label{DetectionDelay}
{\em By selecting $\bar{\xi} = \frac{1}{\sqrt{M}\ln(\gamma)}$ and  $\eta_{\rm G} = (1 + \epsilon)\ln(\gamma)$ for any $\epsilon > 0$, we have $\mathbb{E}_{\infty}(T_{\rm G})\geq \gamma$ as $\gamma \rightarrow \infty$. If, in addition, Assumption \ref{AverageAttackPowerAssumption} holds, $\mathcal{D}(T_{\rm G}) =  \mathcal{O}\left( \frac{\ln \gamma}{M( \mu - \ln(1 + \mu) )}\right) $ as $\gamma\rightarrow \infty$.}
\end{theorem}
\begin{IEEEproof}
Please refer to Appendix \ref{App:DetectionPerformance}.
\end{IEEEproof}

Theorem \ref{DetectionDelay} has characterized the performance of $T_G$ in terms of the ARL2FA and the WADD. Based on Theorem \ref{DetectionDelay}, we now analyze $\mathcal{W}(T_{\rm G})$.
According to \eqref{SignalReceivedbyEve}, under the condition that Eve performs IRS-PCA and Alice uses MRT beamforming, the channel capacity of Eve in $\mathcal{T}_k$  is
\begin{align}
C_{e,k}^{\rm \text{IRS-PCA}} &= \ln\left(1 + {\rm SNR}_{e,k}^{\rm \text{mrt,IRS-PCA}}\right)
\end{align}
where ${\rm SNR}_{e,k}^{\rm \text{mrt,IRS-PCA}} \triangleq \frac{1}{\sigma_e^2} \left|\left( a_e \bm{h}_{e,k}^H + \tilde{a}_{{\rm I},k} \bm{h}_{{\rm I},k}^H\right)\bm{w}_{\rm mrt}\right|^2$.
The following theorem gives a upper bounded on the mean value of $C_{e,k}^{\rm \text{IRS-PCA}}$.
\begin{theorem}
\label{Theorem:UpperBoundOnAttackingCapacity}
{\em Given $ \tilde{a}_{{\rm I},k}$ and $ \hat{a}_{{\rm I},k}$, an upper bound on $\mathbb{E}\{C_{e,k}^{\rm IRS\text{-}PCA}\}$ is
\begin{align}
\label{NoCountermeasureWiretappingRate}
&\quad\,\, \mathbb{E}\{C_{e,k}^{\rm IRS\text{-}PCA}\} \nonumber \\
&\leq \ln \left(1 + \frac{a_e^2}{\sigma_e^2}
+
\frac{\left| \tilde{a}_{{\rm I},k} \right|^2 }{\sigma_e^2}
\frac{ \mu_k }{ 1 + \mu_k  }M
\left(1 + o_M(1)\right)\right),
\end{align}
where a detailed expression for $o_M(1)$ is given in \eqref{UpperBoundDerivation}.}
\end{theorem}
\begin{IEEEproof}
Suppose $\bm{x}\sim\mathcal{CN}(\bm{0},\bm{I}_M)$, $\bm{y}\sim\mathcal{CN}(\bm{0},\bm{I}_M)$, $\bm{z}\sim\mathcal{CN}(\bm{0},\bm{I}_M)$ and $\bm{x},\bm{y},\bm{z}$ are mutually independent.
We have
\begin{align}
\mathbb{E}\left\{ {\rm SNR}_{e,k}^{\rm \text{mrt,IRS-PCA}} \right\}& = \mathbb{E}\left\{
\frac{\left|( \tilde{a}_{{\rm I},k} \bm{x} + a_e \bm{y})^H ( \hat{a}_{{\rm I},k} \bm{x} + \sigma_0\bm{z})\right|^2 }
{\sigma_e^2\|\hat{a}_{{\rm I},k} \bm{x} + \sigma_0\bm{z}\|^2}\right\}\nonumber \\
&= \frac{a_e^2}{\sigma_e^2}  + \frac{|\tilde{a}_{{\rm I},k}|^2}{\sigma_e^2} \Pi,
\end{align}
where $\Pi \triangleq \mathbb{E}\left\{
\frac{\left|\bm{x}^H ( \hat{a}_{{\rm I},k} \bm{x} + \sigma_0\bm{z})\right|^2 }
{ \|\hat{a}_{{\rm I},k} \bm{x} + \sigma_0\bm{z}\|^2}\right\} $. An upper bound on $\Pi$ is derived as follows,
\begin{align}
\Pi &= \mathbb{E}\left\{
\frac{\left|\bm{x}^H \left( \left(\hat{a}_{{\rm I},k} + \sigma_0\frac{\chi}{\|\bm{x}\|}\right) \bm{x} + \sigma_0\bm{z}^{\prime}\right)\right|^2 }
{ \left\|\left(\hat{a}_{{\rm I},k} + \sigma_0\frac{\chi}{\|\bm{x}\|}\right) \bm{x} + \sigma_0\bm{z}^{\prime}\right\|^2}\right\}\nonumber\\
&=  \mathbb{E}\left\{
\frac{  \left| \hat{a}_{{\rm I},k} + \sigma_0\frac{\chi}{\|\bm{x}\|} \right|^2 \|\bm{x}\|^4 }
{\left| \hat{a}_{{\rm I},k} + \sigma_0\frac{\chi}{\|\bm{x}\|}\right|^2 \|\bm{x}\|^2 + \sigma_0^2 \|\bm{z}^{\prime}\|^2}\right\}
 \nonumber\\
&\leq \mathbb{E}\left\{
\frac{ \left| \hat{a}_{{\rm I},k} \right|^2\|\bm{x}\|^4 + \sigma_0^2  \|\bm{x}\|^2 }
{ \left| \hat{a}_{{\rm I},k} \right|^2\|\bm{x}\|^2 + \sigma_0^2  + \sigma_0^2 \|\bm{z}^{\prime}\|^2}\right\}\nonumber \\
&\leq \mathbb{E}\left\{
\frac{ \left| \hat{a}_{{\rm I},k} \right|^2\|\bm{x}\|^4 + \sigma_0^2  \|\bm{x}\|^2 }
{ \left| \hat{a}_{{\rm I},k} \right|^2\|\bm{x}\|^2 +  \sigma_0^2(M-1)}\right\} \nonumber\\
&=  M \mathbb{E}\left\{
\frac{\left| \hat{a}_{{\rm I},k} \right|^2\mathcal{G}(M+1,1) + \sigma_0^2   }
{ \left| \hat{a}_{{\rm I},k} \right|^2\mathcal{G}(M+1,1) + \sigma_0^2(M-1)}\right\}\nonumber \\
& \leq  M
\frac{\left| \hat{a}_{{\rm I},k} \right|^2(M+1)  + \sigma_0^2 }
{ \left| \hat{a}_{{\rm I},k} \right|^2(M+1)  +  \sigma_0^2(M-1)} \nonumber\\
&
=  \frac{ \left| \hat{a}_{{\rm I},k} \right|^2 }{ \left| \hat{a}_{{\rm I},k} \right|^2 + \sigma_0^2 }M
\Bigg(1 + \frac{
\left(\frac{1}{\psi_{3,k}} - \psi_{3,k} + \psi_{4,k}\right)\frac{1}{M}}
{ 1 + (\psi_{3,k} - \psi_{4,k})\frac{1}{M} } \Bigg)\nonumber \\
&= \frac{ \left| \hat{a}_{{\rm I},k} \right|^2 }{ \left| \hat{a}_{{\rm I},k} \right|^2 + \sigma_0^2 }M
\Bigg(1 + o_M(1)\Bigg),
\label{UpperBoundDerivation}
\end{align}
where
$\chi = \frac{\bm{x}^H\bm{z}}{\|\bm{x}\|}\sim\mathcal{CN}(0,1)$,
$\bm{z}^{\prime} = \left(\bm{I}_M - \frac{\bm{x}\bm{x}^H}{\|\bm{x}\|^2}\right)\bm{z}$,
$\psi_{3,k} = \frac{\left| \hat{a}_{{\rm I},k} \right|^2}{\left| \hat{a}_{{\rm I},k} \right|^2 + \sigma_0^2}$,
$\psi_{4,k} = \frac{\sigma_0^2}{\left| \hat{a}_{{\rm I},k} \right|^2 + \sigma_0^2}$,
and the derivation follows the fact that $\frac{x+b}{x+a}$ is concave in $x\in(0,\infty)$ if $a>b\geq 0$,  $\mathbb{E}\left\{\frac{1}{\mathcal{G}(M,1)}\right\} = \frac{1}{M-1}$, and $\|\bm{z}^{\prime}\|^2\sim\mathcal{G}(M-1,1)$ is independent of $\chi$.
\end{IEEEproof}

\begin{corollary}
\label{Coro:AWTMassiveMIMORegion}
{\em  Given $ \tilde{a}_{{\rm I},k}$ and $ \hat{a}_{{\rm I},k}$, an upper bound on $\mathbb{E}\left\{\Delta\mathcal{I}_k\right\}$ is
\begin{align}
\mathbb{E}\left\{\Delta\mathcal{I}_k\right\}\leq \tau_d \Bigg( &\ln\left( 1 + \frac{a_e^2}{\sigma_e^2} +
\frac{\left| \tilde{a}_{{\rm I},k} \right|^2}{\sigma_e^2} \frac{\mu_k M}{\mu_k + 1} (1 + o_M(1)) \right) \nonumber \\
&- \mathrm{e}^{\frac{\sigma_e^2}{a_e^2}}
\mathrm{E}\left( \frac{\sigma_e^2}{a_e^2}\right) \Bigg).
\end{align}
Furthermore, for massive MIMO systems, i.e., $M$ is sufficiently large,  we have $\mathbb{E}\left\{\Delta\mathcal{I}_k\right\} \leq \mathcal{O}\left(\ln M\right)$.
For the case where Eve works in a low SNR region,
$\mathbb{E}\left\{\Delta\mathcal{I}_k\right\} \leq \mathcal{O}\left(\frac{\left| \tilde{a}_{{\rm I},k} \right|^2}{\sigma_e^2} \frac{\mu_k}{\mu_k + 1}M\right)$.
}
\end{corollary}
\begin{IEEEproof}
Due to the fact $\frac{ \left|\bm{h}_{e,k}^H\left(\bm{h}_{b,k} + \bm{z}_k\right)\right|^2}{\left\|\bm{h}_{b,k} + \bm{z}_k\right\|^2}\sim\mathcal{E}(1)$, we obtain that
\begin{align}
\label{NoIRSCapacity}
\mathbb{E}(C_{e,k}^{\rm \text{no~IRS}}) &= \int_0^\infty\mathrm{e}^{-x}\ln\left(1 + \frac{a_e^2}{\sigma_e^2}x\right)\mathrm{d}x\nonumber \\
&= \mathrm{e}^{\frac{\sigma_e^2}{a_e^2}}
\mathrm{E}\left( \frac{\sigma_e^2}{a_e^2}\right) \sim \frac{a_e^2}{\sigma_e^2},
\end{align}
where $\mathrm{E}(x) \triangleq \int_1^\infty \frac{\mathrm{e}^{-tx}}{t}\mathrm{d}t$, and  ``$\sim$'' holds when $\frac{a_e^2}{\sigma_e^2}\rightarrow 0$.
Combining \eqref{NoIRSCapacity} with Theorem \ref{Theorem:UpperBoundOnAttackingCapacity} leads to this corollary.
\end{IEEEproof}
\begin{corollary}
\label{AWTMassiveMIMORegion}
{\em Suppose $\bm{\Phi}^{(d)} = \bm{\Phi}_{\nu+k}^{(d)}$ and $\bm{\Phi}^{(p)} = \bm{\Phi}_{\nu+k}^{(p)}$ for $\forall k \geq 0$. For massive MIMO system, i.e., $M$ is sufficiently large, $\mathcal{W}(T_{\rm G})$ is upper bounded by $\mathcal{O}\left(  \frac{\ln M}{M} \frac{\tau_d \ln \gamma}{ \mu - \ln(1 + \mu) } \right)$ as $\gamma\rightarrow \infty$.
}
\end{corollary}

\begin{corollary}
\label{AWTLowSNRRegion}
{\em Suppose $\bm{\Phi}^{(d)} = \bm{\Phi}_{\nu+k}^{(d)}$ and $\bm{\Phi}^{(p)} = \bm{\Phi}_{\nu+k}^{(p)}$ for $\forall k \geq 0$. If Eve works in a low SNR region, $\mathcal{W}(T_{\rm G})$ is upper bounded by $
\mathcal{O} \left(N^2  \frac{\mu}{\mu + 1}  \frac{\tau_d \ln \gamma}{ \mu - \ln(1 + \mu) }\right)$
as $\gamma\rightarrow \infty$.}
\end{corollary}
\begin{IEEEproof}
Please refer to Appendix \ref{Appendix:WAWTG}.
\end{IEEEproof}

Corollary \ref{AWTMassiveMIMORegion} reveals that $\mathcal{W}(T_{\rm G})$ decreases with $M$ when $M$ is large.
This is because the detection delay decreases with $M$, and if $M$ is sufficiently large, Alice is able to immediate discover the occurrence of IRS-PCA.
Corollary \ref{AWTLowSNRRegion} reveals that under the condition that the SNR of Eve is low, $\mathcal{W}(T_{\rm G})$ is nearly independent of the number of antennas of Alice $M$. This is because in this case, $\mathbb{E}\left\{\Delta\mathcal{I}_k\right\}$ is linearly to $M$, and the detection delay is inversely proportional to $M$ as shown in Theorem \ref{DetectionDelay}, the effects of which cancel out. However, Corollary \ref{AWTLowSNRRegion} also reveals that in this case $\mathcal{W}(T_{\rm G}) \propto N^2$, and thus Eve can significantly increase its wiretapping capability by increasing the number of reflecting elements of its IRS.

In this section, we have proposed a sequential detection scheme, namely the GCUSUM scheme, for Alice to detect IRS-PCA.
In next section, under the condition that Alice has confirmed the existence of IRS-PCA, we introduce a channel estimation method for Alice to estimate the channel of the IRS, which can be used to reduce wiretapping SNR during the DT phase.

\section{A Cooperative Channel Estimation and Beamforming Scheme}
\label{Sec:SecureTransmission}
In this section, we present our channel estimation scheme for Alice to estimate the channel of the IRS under the condition that Eve performs IRS-PCA.
We introduce several CNs to assist the channel estimation process of Alice.
During the RPT phase, we let each of the CNs broadcast a signal sequence that is orthogonal to the pilot sequence sent by Bob.
Then, except for the pilot sequence sent by Bob, the signal sequences sent by the CNs will also be reflected by the IRS.
By matching Alice's received signal matrix with each of the signal sequences sent by the CNs, Alice obtains multiple sample observations on the channel of the IRS, based on which maximal likelihood channel estimator can be constructed to estimate the channel of the IRS.

Denote $J$ as the number of the CNs in the system. During the RPT phase, the $j$-th CN transmits  $\bm{v}_j\in\mathcal{C}^{\tau_p\times 1}$ to Alice. For $1\leq j\leq J$, we assume that $\|\bm{v}_j\|^2 = \tau_p$ and $\bm{v}_j^H\bm{u} = 0$.
Besides, we assume that $\bm{v}_{1},\bm{v}_{2},\cdots,\bm{v}_{J}$ are mutually orthogonal, namely, $\bm{v}_j^H\bm{v}_i = 0$ for $\forall i\neq j$.
In the following, for notational simplicity, we neglect the subscript $k$ in \eqref{IRSPCAAliceReceives}.
During the RPT phase, the signal matrix received by Alice is
\begin{align}
\bm{Y} =
&\left( a_b \bm{h}_b + a_{{\rm I}} \bm{h}_{\rm I} \right) \bm{u}^H\nonumber \\
&+
\sum_{j=1}^J\left( \sqrt{P_j g_j}  \bm{f}_{j} + \sqrt{P_j g_{\rm I}} \bm{H}_{{\rm I} } \bm{\Phi}^{(p)} \bm{h}_{j,{\rm I}}\right)\bm{v}_j^H +  \bm{Z},
\label{ReceivedPilot}
\end{align}
where $P_j$ is the transmit power of the $j$-th CN, and $g_j$ and $\bm{f}_{j}\sim\mathcal{CN}(\bm{0},\bm{I}_M)$ are the path-loss and fading channel vector of the $j$-th CN, respectively, and $\bm{h}_{j,{\rm I}}$ is the channel vector between the $j$-th CN and the IRS. By matching $\bm{Y}$ with $\bm{v}_j$, Alice obtains
\begin{align}
\bm{t}_j &\triangleq \frac{\bm{Y}\bm{v}_j}{\tau_p \sqrt{P_j g_j}}
= \bm{f}_{j} +  a_j \bm{h}_{\rm I} + \tilde{\bm{z}}_j\nonumber \\
& = \bm{f}_{j} +  b_j \bar{\bm{h}}_{\rm I} + \tilde{\bm{z}}_j
,\quad 1\leq j\leq J,
\label{EstimateHelperChannel}
\end{align}
where $a_j \triangleq \frac{\sqrt{P_j g_{\rm I}}  \bm{\omega}_a^H\bm{\Phi}^{(p)} \bm{h}_{j,{\rm I}}}{\tau_p \sqrt{P_j g_j}}$,
$\tilde{\bm{z}}_j \triangleq \frac{\bm{Z}\bm{v}_j}{\tau_p \sqrt{P_j g_j}}\sim\mathcal{CN}\left(\bm{0},\frac{\sigma_a^2}{\tau_p P_{j}g_j}\bm{I}_M\right)$,
$\bar{\bm{h}}_{\rm I} \triangleq  \frac{\bm{h}_{\rm I}}{ ||\bm{h}_{\rm I}||}$, and
$ b_j \triangleq ||\bm{h}_{\rm I}|| a_j $.

We now derive a maximal likelihood estimation of $\bar{\bm{h}}_{\rm I}$. Based on \eqref{EstimateHelperChannel}, given $\bar{\bm{h}}_{\rm I}$ and $\{b_j : 1\leq j\leq J\}$, the log-likelihood function of $\left(\bm{t}_1,\bm{t}_2,\cdots,\bm{t}_J\right)$ is
\begin{align}
&\quad \ln f\left(\bm{t}_1,\bm{t}_2,\cdots,\bm{t}_J| b_1,b_2,\cdots,b_J,\bar{\bm{h}}_{\rm I}\right)\nonumber \\
&= -\sum_{j=1}^J \frac{\left\|\bm{t}_{j} - b_j \bar{\bm{h}}_{\rm I}\right\|^2}{\sigma_{j}^2} + C,
\end{align}
where $\sigma_{j}^2 \triangleq 1 + \frac{\sigma_a^2}{\tau_p P_{j}g_j}$ for $1\leq j\leq J$, and $C$ is a constant that does not depend on $\{b_j : 1\leq j\leq J\}$ and $\bar{\bm{h}}_{\rm I}$. Then, the maximal likelihood estimation of $\bar{\bm{h}}_{\rm I}$, denoted by $\hat{\bar{\bm{h}}}_{\rm I}$, is
\begin{align}
\hat{\bar{\bm{h}}}_{\rm I}
&= \mathop{\mathrm{argmin}}\limits_{\bm{x};\|\bm{x}\| = 1} \left\{ \min_{b_1,b_2,\cdots,b_j} \sum_{j=1}^J \frac{\left\|\bm{t}_{j} - b_j \bm{x}\right\|^2}{\sigma_{j}^2} \right\}\nonumber \\
&= \mathop{\mathrm{argmin}}\limits_{\bm{x};\|\bm{x}\| = 1}
 \sum_{j=1}^J \frac{\left\|\bm{t}_{j}\right\|^2 - \left|\bm{x}^H\bm{t}_j\right|^2}{\sigma_{j}^2} \nonumber \\
&= \mathop{\mathrm{argmax}}\limits_{\bm{x};\|\bm{x}\| = 1}
 \sum_{j=1}^J \frac{\left|\bm{x}^H\bm{t}_j\right|^2}{\sigma_{j}^2}\nonumber \\
&= \mathop{\mathrm{argmax}}\limits_{\bm{x};\|\bm{x}\| = 1}
 \bm{x}^H\left(\sum_{j=1}^J \frac{1}{\sigma_{j}^2}\bm{t}_j\bm{t}_j^H\right)\bm{x}
=\bm{\lambda}_{\max} \left(\bm{T}\bm{T}^H\right),
\label{ChannelEstimationofIRS}
\end{align}
where $\bm{T}\triangleq \left[ \frac{1}{\sigma_1}\bm{t}_1,\frac{1}{\sigma_2}\bm{t}_2,\cdots, \frac{1}{\sigma_J}\bm{t}_J \right]$, and $\bm{\lambda}_{\max}(\bm{X})$ denotes the normalized eigenvector that corresponds to the largest eigenvalue of $\bm{X}$.
Define $\bm{a}_{\rm CN} \triangleq \left[\frac{a_1}{\sigma_{1} },\frac{a_2}{\sigma_{2} },\cdots,\frac{a_J}{\sigma_{J}}\right]^T$. The following theorem characterizes the accuracy of $\hat{\bar{\bm{h}}}_{\rm I}$ in term of estimating $\bar{\bm{h}}_{\rm I}$.
\begin{theorem}
\label{PropAccuracyOfChannelEstimation}
{\em As $M\rightarrow\infty$, $\hat{\bar{\bm{h}}}_{\rm I} \rightarrow \frac{\bm{h}_{\rm I} + \bm{e}}{\|\bm{h}_{\rm I} + \bm{e}\|}$ , where $\bm{e} \sim\mathcal{CN}\left(\bm{0}, \frac{1}{\|\bm{a}_{\rm CN}\|^2}\bm{I}_M\right)$ is independent of $\bm{h}_{\rm I}$.}
\end{theorem}
\begin{IEEEproof}
Please refer to Appendix \ref{Appendix:EstimationAccuracy}.
\end{IEEEproof}

By \eqref{ReceivedPilot} and \eqref{EstimateHelperChannel},
$\|\bm{a}_{\rm CN}\|^2$ can essentially be understood as the weighted sum power of the signals reflected by the IRS.
Theorem \ref{PropAccuracyOfChannelEstimation} reveals that the estimation error is inversely proportional to $\|\bm{a}_{\rm CN}\|^2$.
In principle, if we increase the number of the CNs, then $\|\bm{a}_{\rm CN}\|^2$ will also be increased, meaning that a more accurate estimation of $\bar{\bm{h}}_{\rm I}$ can be obtained.
Based on \eqref{ChannelEstimationofIRS}, ZF beamforming can be used to facilitate secure transmission in the DT phase.
The ZF beamforming vector is
\begin{align}
\bm{w}_{\rm zf} \triangleq \left(\bm{I}_M - \hat{\bar{\bm{h}}}_{\rm I}\hat{\bar{\bm{h}}}_{\rm I}^H\right)\bm{y} \Big/ \left\|\left(\bm{I}_M - \hat{\bar{\bm{h}}}_{\rm I}\hat{\bar{\bm{h}}}_{\rm I}^H\right)\bm{y}\right\|,
\label{ProposedZFBeam}
\end{align}
where $\bm{y}$ is given by \eqref{BasicIRSPCAQDModelUnderAttack}. Therefore, the wiretapping SNR can be written as
\begin{align}
\mathrm{SNR}_e^{\rm \text{zf,IRS-PCA}} = \frac{1}{\sigma_e^2}\left| \left( a_e \bm{h}_{e}^H + \tilde{a}_{{\rm I}} \bm{h}_{{\rm I}}^H\right) \bm{w}_{\rm zf} \right|^2.
\end{align}
The following theorem characterizes $\mathrm{SNR}_e^{\rm \text{zf,IRS-PCA}}$ in the massive MIMO region.
\begin{theorem}
\label{Therorem:SNREAfterZF}
{\em As $M\rightarrow\infty$, the wiretapping SNR satisfies
\begin{align}
\label{CombatingWiretappingSNR}
&\mathbb{E}\{\mathrm{SNR}_e^{\rm zf,IRS\text{-}PCA}\} \nonumber \\
\rightarrow&
\frac{|\tilde{a}_{{\rm I}} |^2}{\sigma_e^2}
\frac{ (|\hat{a}_{{\rm I}}|^2/\sigma_0^2) M }
{\left(1 + \|\bm{a}_{\rm CN}\|^2\right)\left( (1 + \|\bm{a}_{\rm CN}\|^2) + (|\hat{a}_{{\rm I}}|^2/\sigma_0^2)\right) } + \frac{a_e^2}{\sigma_e^2}
\end{align}
}
\end{theorem}
\begin{IEEEproof}
Please refer to Appendix \ref{Appendix:ProofSNREAfterZF}.
\end{IEEEproof}

By comparing \eqref{CombatingWiretappingSNR} with \eqref{UpperBoundDerivation},
we can see that if $\|\bm{a}_{\rm CN}\|^2$ is sufficiently large, then the wiretapping SNR can be significantly reduced with the aided of the propose ZF beamforming scheme.
This is the direct result of the fact that in this case, the channel of the IRS can be accurately estimated, as shown in Theorem \ref{PropAccuracyOfChannelEstimation}

\begin{remark}
It should be noted that the CNs play an important role in the proposed cooperative channel estimation scheme.
Specifically, each CN produces an independent sample observations on $\bar{\bm{h}}_{\rm I}$ to facilitate Alice to estimate $\bar{\bm{h}}_{\rm I}$, as shown in \eqref{EstimateHelperChannel}.
In practice, the CNs can be some secure anchor nodes that are under the control of the legitimate system or other legitimate user devices in the considered system.
For example, in cellular networks, the CNs can be other mobile users that are not scheduled in the considered time blocks.
Another example is that in wireless sensor networks, if a sensor node intends to communication with the fusion center, then other sensor nodes can act as cooperative nodes.
Note that it is possible that there is no CN in the system, and in this case, Alice can only estimate the channels of Bob and the IRS by using the signal model in \eqref{BasicIRSPCAQDModelUnderAttack}.
Existing methods in \cite{Y.WuTIT2016,H.-MWang2018} facilitate Alice to do so under the condition that the covariance matrices of $\bm{h}_{b}$ and $\bm{h}_{{\rm I}}$ are different and perfectly known to Alice.
Different from the methods in \cite{Y.WuTIT2016,H.-MWang2018},
the proposed method exploits the CNs in the considered system to estimate $\bar{\bm{h}}_{\rm I}$, and does not require any knowledge about the distribution of $\bar{\bm{h}}_{\rm I}$.
\end{remark}

\section{Numerical Results}
\label{Sec:NumericalSection}

In this section, we present some simulation results to gain some insights about the problem of IRS-PCA investigated in previous sections.

\begin{figure*}[t]
  \centering
  \begin{minipage}[t]{0.48\linewidth}
  \centering
  \includegraphics[width=2.7 in]{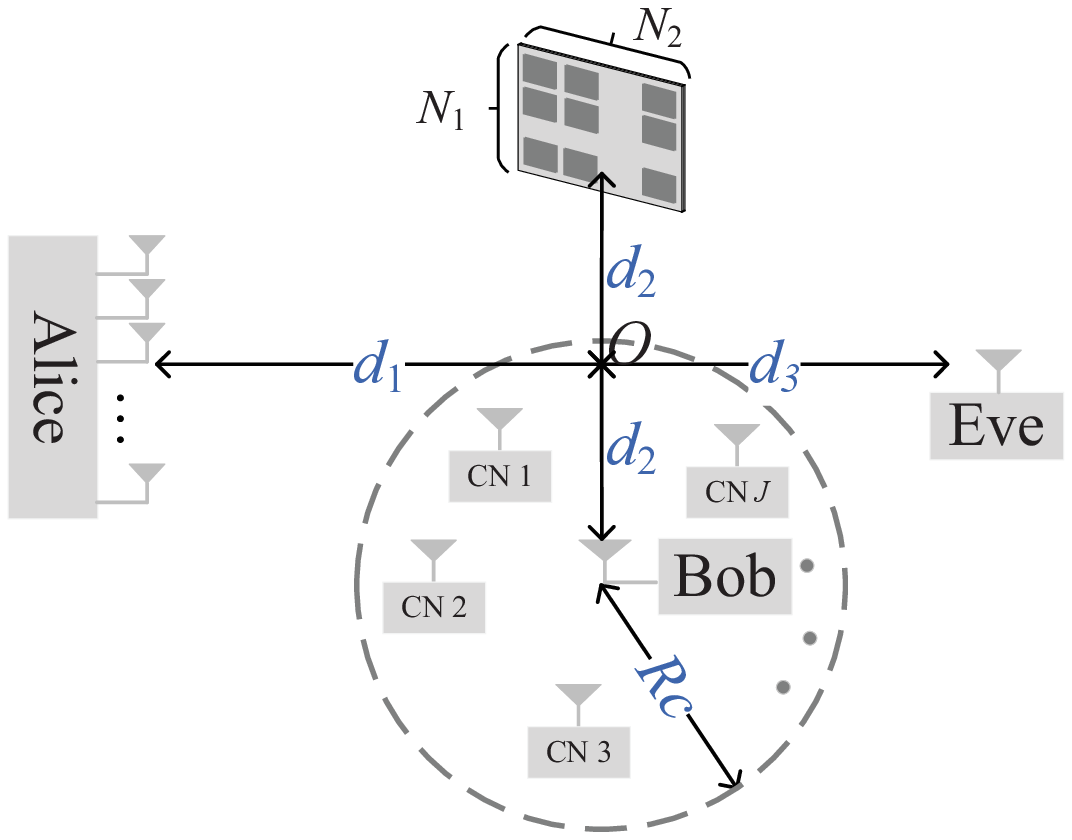}
  \caption{\small Basic settings in simulation.}\label{SimulationSetting}
  \end{minipage}
  \hspace{0.1in}
  \begin{minipage}[t]{0.48\linewidth}
  \centering
  \includegraphics[width=2.7 in]{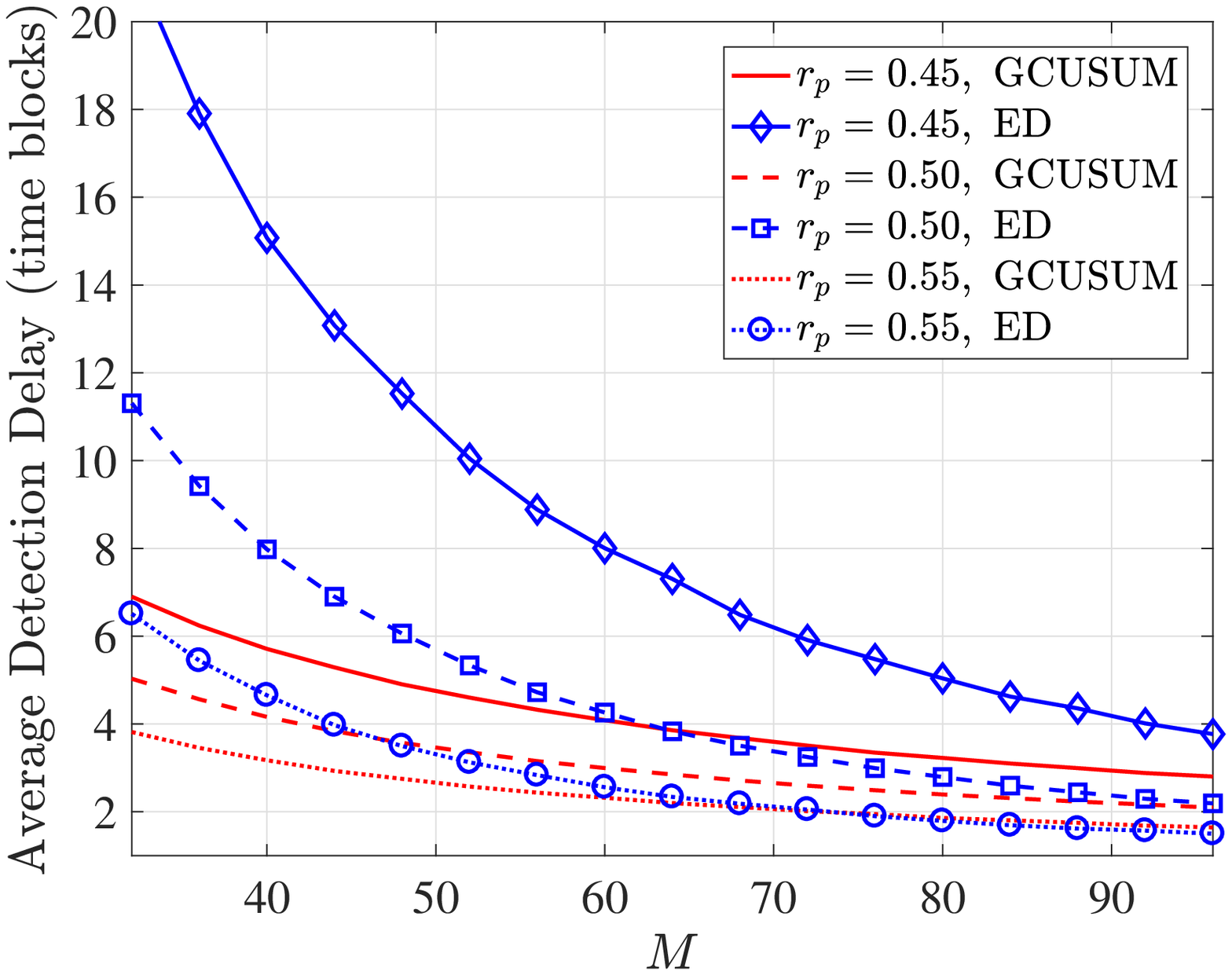}
  \caption{\small Average detection delay versus $M$.}\label{ADDvsM}
  \end{minipage}%
  \vspace{-3mm}
\end{figure*}

We illustrate our simulation settings in Fig. \ref{SimulationSetting}. Suppose that Alice, Bob, Eve, and the IRS are in the same plane and their locations are $(-d_1,0)$, $(0,-d_2)$, $(d_3,0)$, and $(0,d_2)$, respectively.
For the proposed cooperative channel estimation scheme, we randomly generate $J$ CNs in a circular area with the center and the radius being $(0,-d_2)$ and $R_c$, respectively.
The distances from Alice to Bob, Eve, the IRS, and the $j$-th CN are denoted by
$d_{b,a}$, $d_{e,a}$, $d_{{\rm I},a}$, and $d_{j,a}$, respectively. The distances from the IRS to Bob, Eve, and the $j$-th CN are denoted by $d_{b,{\rm I}}$, $d_{e,{\rm I}}$, and $d_{j,{\rm I}}$, respectively.
The path losses are set to be
$g_b=d_{b,a}^{-4}$,
$g_e=d_{e,a}^{-4}$,
$g_{\rm I}=d_{{\rm I},a}^{-4}$, and
$g_j=d_{j,a}^{-4}$.
The channels from Bob, Eve, and the $j$-th CN to the IRS are assumed to AWGN channels with
$\bm{h}_{b,{\rm I},k} = d_{b,{\rm I}}^{-2}\bm{\omega}_b$, $\bm{h}_{e,{\rm I},k} = d_{e,{\rm I}}^{-2}\bm{\omega}_e$, and $\bm{h}_{j, {\rm I}} = h_{j,{\rm I}}\bm{\omega}_j$ with $h_{j,{\rm I}} = d_{j,{\rm I}}^{-2}$. The IRS of Eve consists of $N = N_1\times N_2$ reflecting elements which form a $N_1$-by-$N_2$ planar array. Accordingly, we set $\bm{\omega}_a$, $\bm{\omega}_b$, $\bm{\omega}_e$, and $\bm{\omega}_j$ to be the steering vectors of such a planar array with the angle of arrivals depending on the locations of these nodes as introduced. For the passive beamforming at the IRS, we set $\bm{\Phi}_k^{(p)} = r_p\times{\rm diag}\{\bm{\omega}_a \}\times {\rm diag}\{\bm{\omega}_b^*\}$ and
$\bm{\Phi}_k^{(d)} = r_d\times{\rm diag}\{\bm{\omega}_e  \}\times {\rm diag}\{\bm{\omega}_a^*\}$ so that the beams of the IRS during the RPT and DT phases align with the Bob-IRS-Alice and the Alice-IRS-Eve channels, respectively. Unless specified, we set $\nu = 1$, $r_p = r_d = 1$, $d_1 = 150$ m, $d_2 = 20$ m, $d_3 = 30$ m, $R_c = 30$ m, $\tilde{N} = N_1 = N_2 = 7$,  $J = 15$, $P_b = 20$ dBm, $P_{j} = 20$ dBm for $1\leq j\leq J$, $\sigma_a^2 = \sigma_e^2 = \sigma_b^2 = -80$ dBm with $P_a$ normalizing $\frac{P_a g_b}{\sigma_b^2}$ to $0$ dB.

\subsection{IRS-PCA detection}
\label{Sec:NumericalSectionIRS-PCA detection}
\begin{figure*}[t]
  \centering
  \begin{minipage}[t]{0.48\linewidth}
  \centering
  \includegraphics[width=2.8 in]{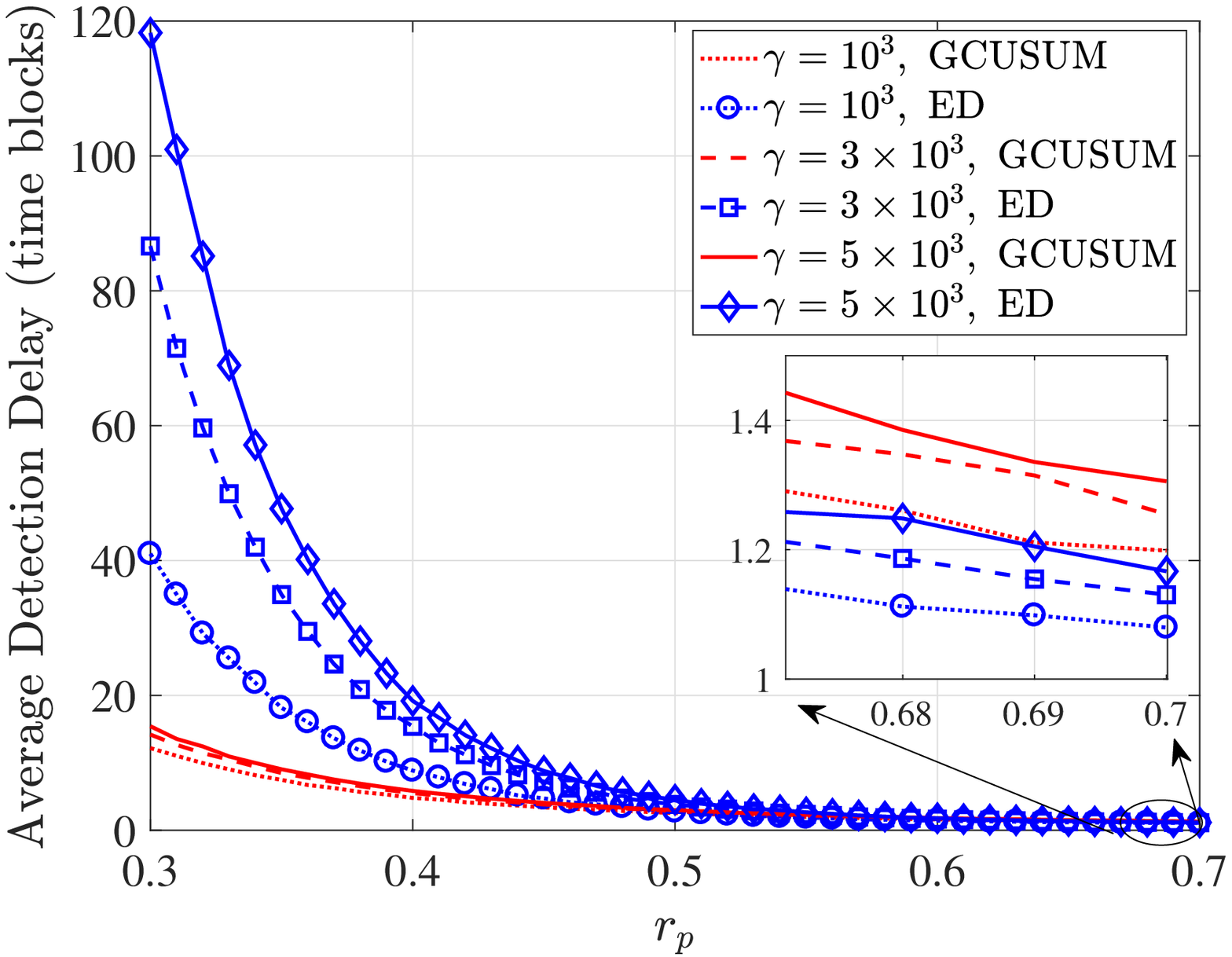}
  \caption{\small Average detection delay versus $r_p$.}\label{ADDvsRp}
  \end{minipage}
  \hspace{0.1in}
  \begin{minipage}[t]{0.48\linewidth}
  \centering
  \includegraphics[width=2.8 in]{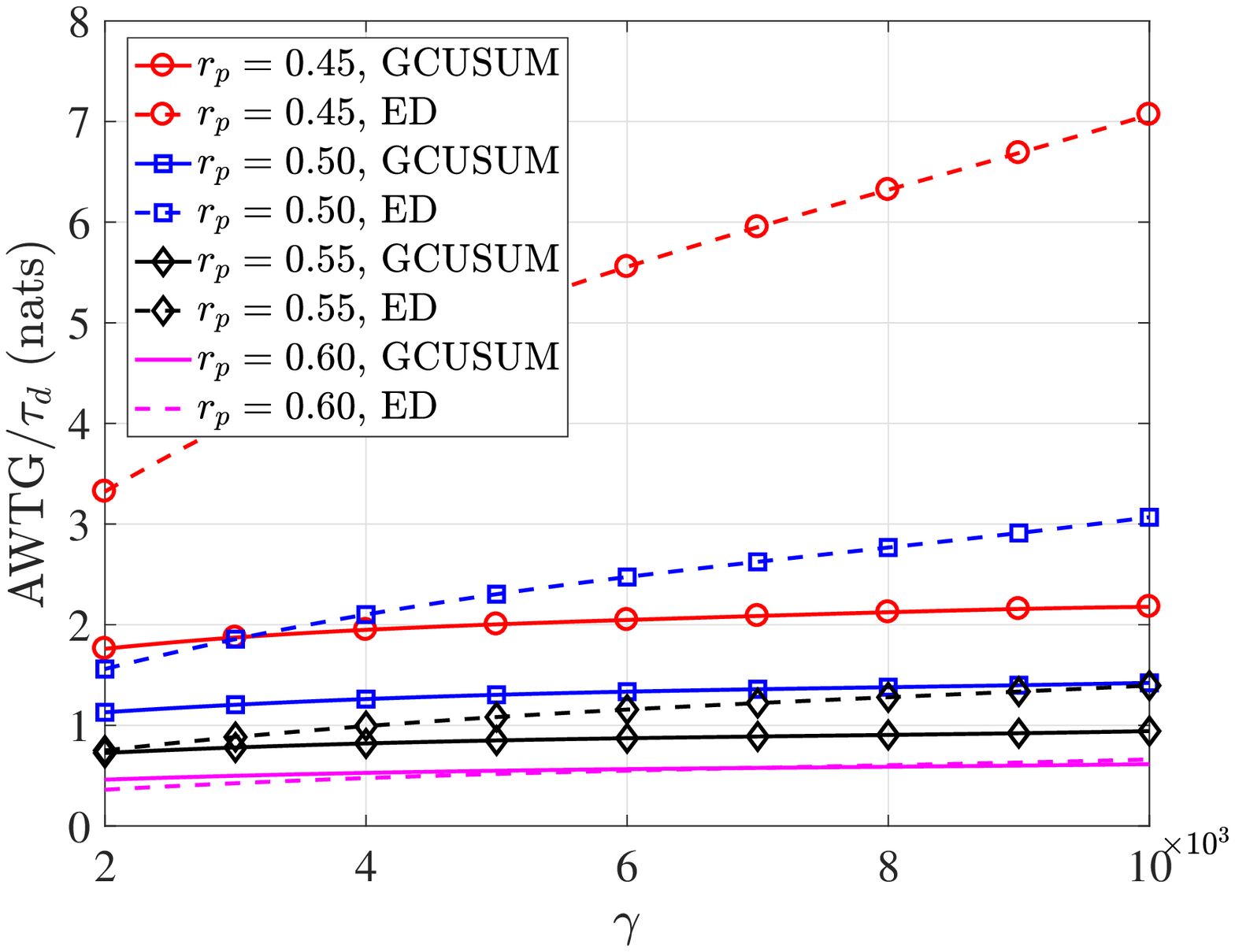}
  \caption{\small $\mathcal{W}(T)$ versus $\gamma$.}\label{WAWTGvsARLFA}
  \end{minipage}%
  \vspace{-3mm}
\end{figure*}

In this subsection, we present numerical results to show the detection performance of the proposed IRS-PCA detection scheme. An energy-based detection (ED) scheme is considered for comparison. In the ED scheme, in the $k$-th time block, Alice compares $\|\bm{y}_k\|^2$ with a pre-designed detection threshold $\eta_{E}$. If $\|\bm{y}_k\|^2 > \eta_{E}$, then the detector raises an alarm. The detection procedure of the ED scheme can be characterized by  a stopping time defined as $T_E = \inf\{k:k\geq 1, \|\bm{y}_k\|^2 > \eta_{E}\}$.
In the simulation, we set the ARL2FA of the ED scheme to be $\gamma$, i.e., $\mathbb{E}_{\infty}(T_E) = \gamma$. Due to the fact that if no IRS-PCA exists, $\frac{\|\bm{y}_k\|^2}{\sigma_0^2}\sim\mathcal{G}(M,1)$, the detection threshold of the ED scheme is given by $\eta_{E} = \sigma_0^2\Gamma_M^{-1}(1/\gamma)$, where $\Gamma_M^{-1}(\cdot)$ is the inverse of the incomplete Gamma function. For the proposed GCUSUM scheme, we set $\bar{\xi} = \frac{1}{\sqrt{M}\ln(\gamma)}$ and $\eta_{G} = \ln(\gamma)$. Extensive numerical experiments reveal that the ARL2FA of the GCUSUM scheme is strictly larger than $\gamma$ for the values of $\gamma$ considered in the simulation.

In Fig. \ref{ADDvsM}, we illustrate the average detection delay versus the antenna number of Alice, where we set $\gamma = 3\times 10^3$.
It can be seen from Fig. \ref{ADDvsM} that both the GCUSUM and the ED schemes present shorter average detection delays as the number of antennas of Alice increases. It is worth noting that when $M$ is not large, the average detection delay of the proposed GCUSUM scheme is much shorter than the ED scheme.
In Fig. \ref{ADDvsRp}, we illustrate the average detection delay versus the amplitude of the reflecting coefficient of the IRS $r_p$, where we set $M = 64$. In general, as $r_p$ increases, the difference between the probability distributions of $\bm{y}_k$ when Eve does not perform IRS-PCA and when IRS-PCA truly occurs becomes more significant, and thus the IRS-PCA will be detected by Alice more quickly.
Fig. \ref{ADDvsRp} reveals that when $r_p$ is small, the performance of the proposed GCUSUM scheme is better than the ED scheme.
If $r_p$ is sufficiently large, both the GCUSUM and the ED schemes exhibit short detection delay, and the ED scheme works slightly better than the proposed GCUSUM scheme.
In Fig. \ref{WAWTGvsARLFA}, we plot the normalized (by the length of the DT phase) WAWTG versus the ARL2FA of Alice's detector, i.e., $\gamma$.
From Fig. \ref{WAWTGvsARLFA}, it can be seen that WAWTG increases with $\gamma$ under both the GCUSUM and the ED schemes, however the increasing rates are quite different. When $r_p$ is small, $\mathcal{W}(T_E)$ increases much faster than $\mathcal{W}(T_G)$, and compared with the GCUSUM scheme, Eve can intercept more data before the IRS-PCA is successfully detected by Alice if Alice adopts the ED scheme. When $r_p$ is large, both the GCUSUM and the ED schemes restrict the WAWTG to be small, but as $\gamma$ increases, $\mathcal{W}(T_E)$ tends to be larger than $\mathcal{W}(T_G)$.

In summary, Fig. \ref{ADDvsM}, \ref{ADDvsRp}, and \ref{WAWTGvsARLFA} reveal that there exists a nontrivial region of $(M,r_p,\gamma)$ wherein the proposed GCUSUM scheme can discover the occurrence of IRS-PCA more quickly and lead to a weaker wiretapping capability of Eve than the benchmark ED scheme.

\subsection{Secure transmission under IRS-PCA}
\begin{figure*}[t]
  \centering
  \begin{minipage}[t]{0.48\linewidth}
  \centering
  \includegraphics[width=2.8 in]{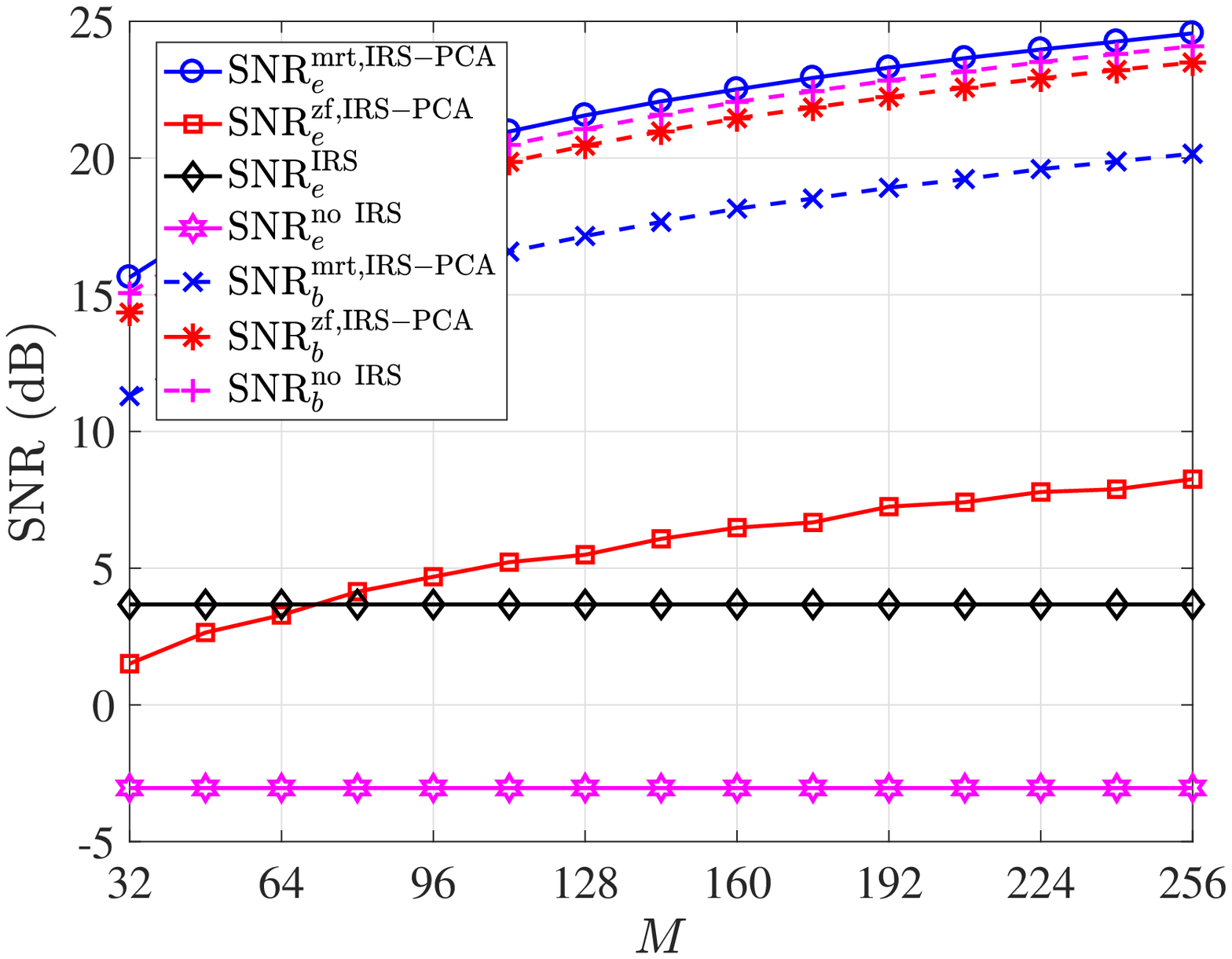}
  \caption{\small SNR versus M.}\label{STSNRvsM}
  \end{minipage}
  \hspace{0.1in}
  \begin{minipage}[t]{0.48\linewidth}
  \centering
  \includegraphics[width=2.8 in]{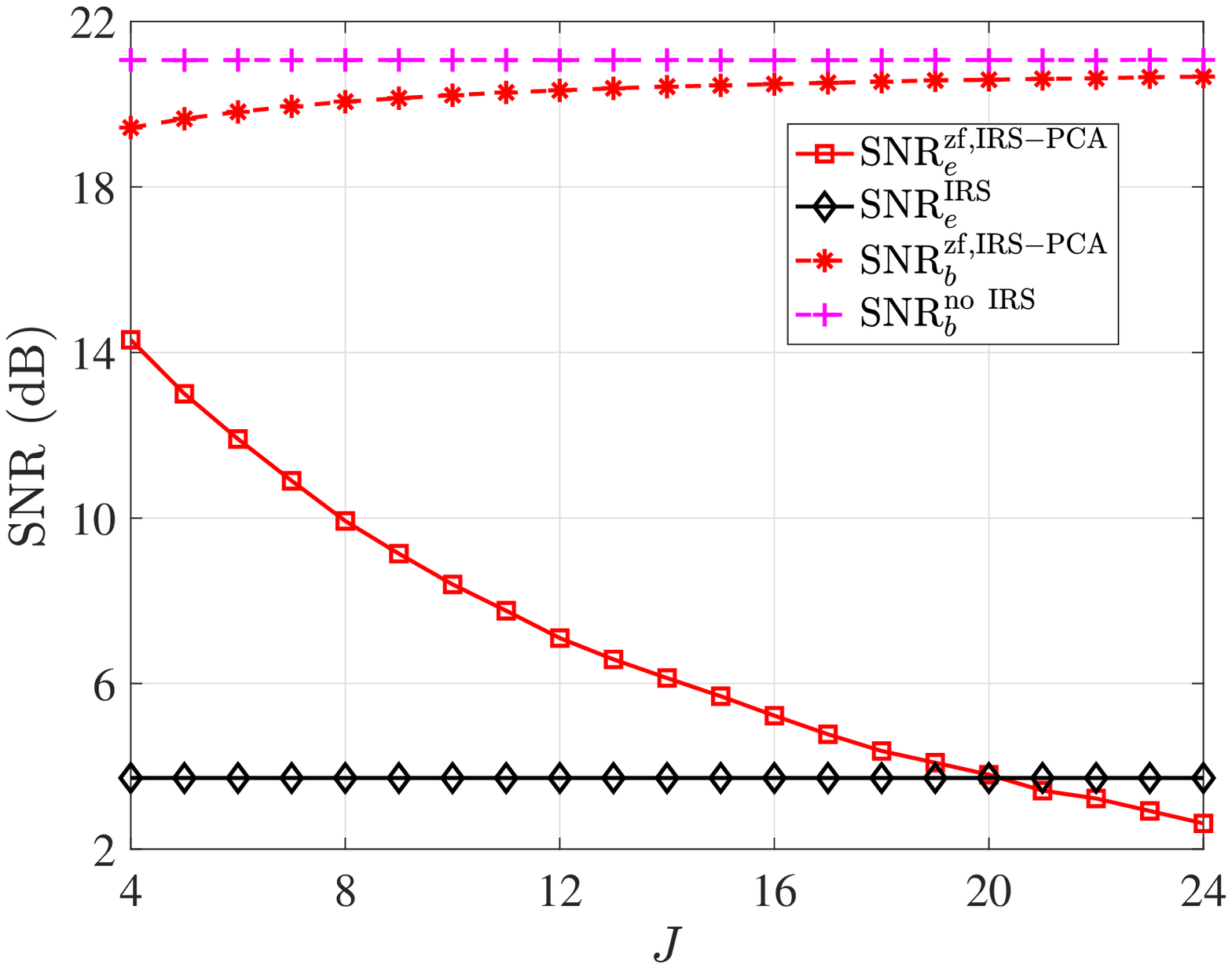}
  \caption{\small SNR versus J.}\label{STSNRvsJ}
  \end{minipage}%
  \vspace{-3mm}
\end{figure*}

\begin{figure*}[t]
  \centering
  \begin{minipage}[t]{0.48\linewidth}
  \centering
  \includegraphics[width=2.8 in]{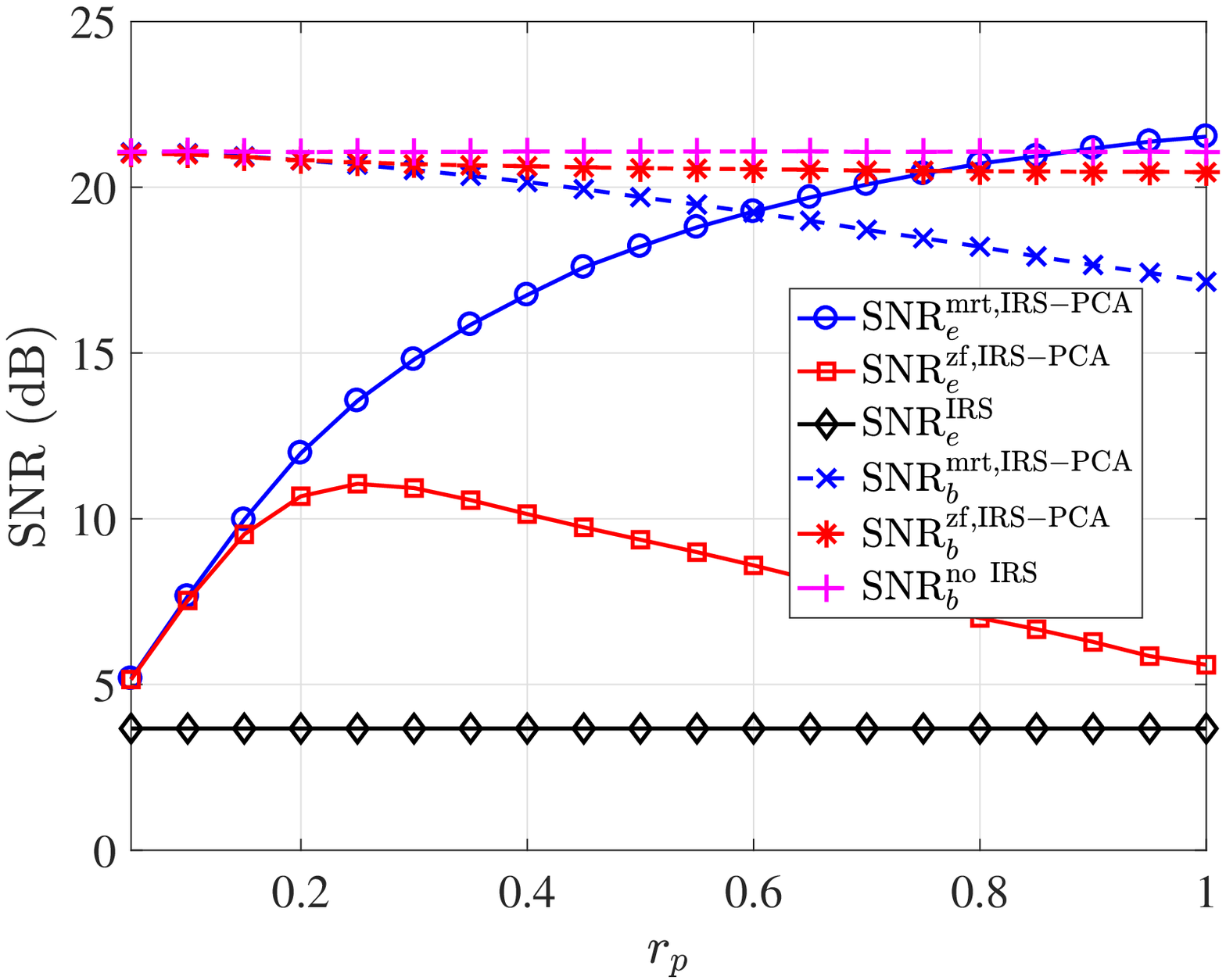}
  \caption{\small SNR versus $r_p$}\label{STSNRvsrp}
  \end{minipage}
  \hspace{0.1in}
  \begin{minipage}[t]{0.48\linewidth}
  \centering
  \includegraphics[width=2.8 in]{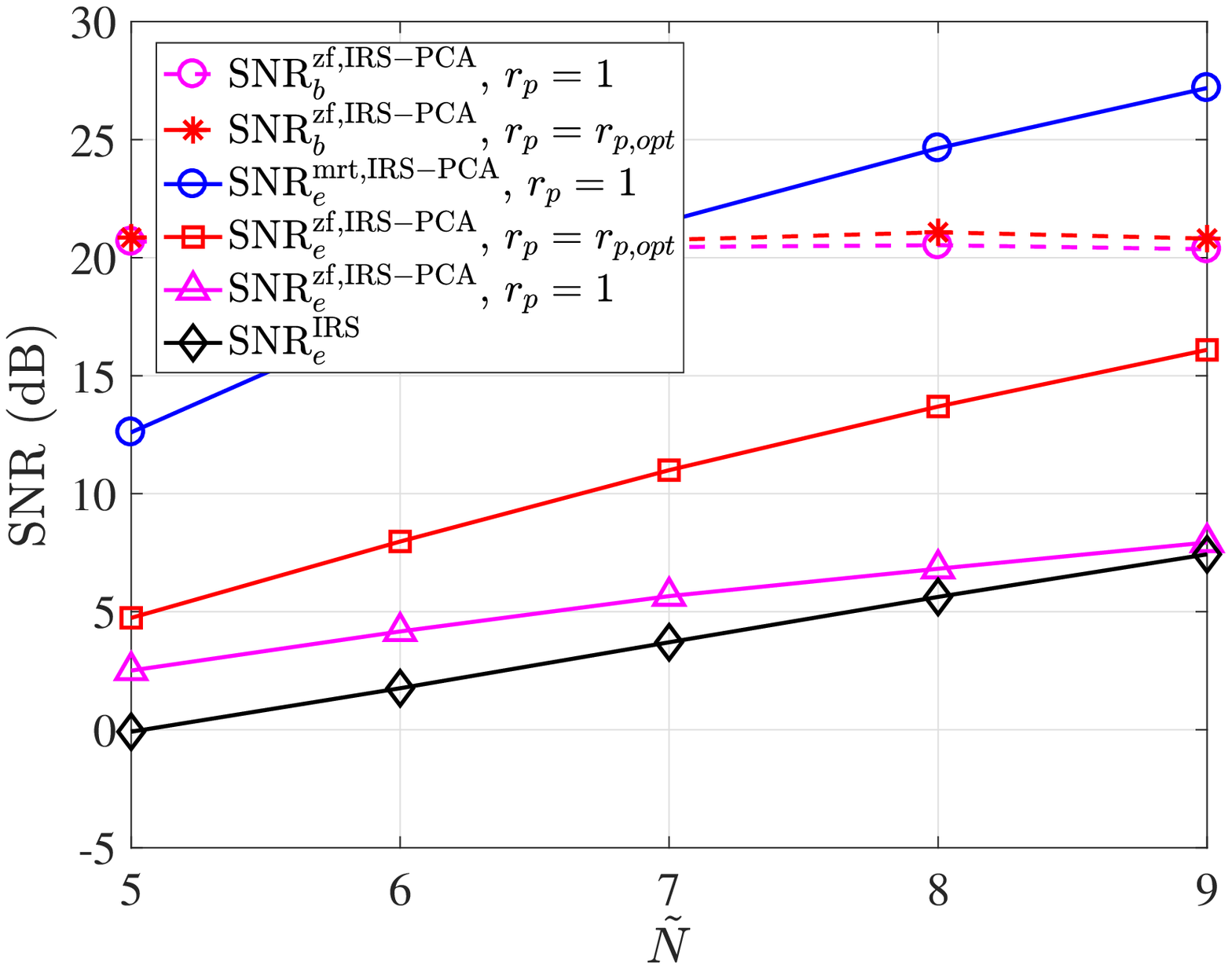}
  \caption{\small SNR versus $\tilde{N}$ where $r_{p,opt}$ denotes the optimal value of $r_p$ that maximizes the SNR of Eve.}\label{STSNRvsNtidle}
  \end{minipage}%
  \vspace{-3mm}
\end{figure*}
In this subsection, numerical results are presented to show the performance of the proposed secure transmission scheme.
We evaluate the SNR of Bob and Eve during the DT phase under the following conditions,
\begin{enumerate}
\item ${\rm SNR}_X^{\text{no IRS}}$: There is no IRS in the system. Eve receives the signal from Alice passively and does not launch any attack;
\item ${\rm SNR}_X^{\text{IRS}}$: Eve uses an IRS to enhance its receiving SNR, but does not perform IRS-PCA, i.e., the IRS is turned off during the RPT phase.
\item ${\rm SNR}_X^{\text{mrt,IRS-PCA}}$: Eve performs IRS-PCA, and Alice uses the naive MRT beamforming scheme  to transmit its data to Bob.
\item ${\rm SNR}_X^{\text{zf,IRS-PCA}}$: Eve performs IRS-PCA, and Alice estimates $\bar{\bm{h}}_{\rm I}$ by using the scheme proposed in Section \ref{Sec:SecureTransmission} and adopts the ZF beamforming scheme in \eqref{ProposedZFBeam}.
\end{enumerate}
where the subscript $X\in\{b,e\}$ indicates Bob or Eve.

In Fig. \ref{STSNRvsM},  we illustrate the SNRs of Bob and Eve versus the antenna number of Alice.
First of all, we point out that when there is no IRS or when Eve only uses the IRS to enhance its receiving signal strength during the DT phase, Alice's beamforming vector is independent of the channels of Eve and the IRS, i.e. $\bm{h}_e$ and $\bm{h}_{\rm I}$, and thus ${\rm SNR}_e^{\text{no IRS}}$ and ${\rm SNR}_e^{\text{IRS}}$ are independent of $M$.
From Fig. \ref{STSNRvsM}, it is worth noting that even if Eve does not perform IRS-PCA, there exhibits a notable increase on its wiretapping SNR due to the use of an IRS.
If, in addition, Eve performs IRS-PCA, it can obtain a huge improvement on its wiretapping SNR (see the curve of ${\rm SNR}_e^{\text{ mrt,IRS-PCA}}$), and in this case, ${\rm SNR}_e^{\text{ mrt,IRS-PCA}}$ becomes increasing with $M$.
Besides, the curves of ${\rm SNR}_b^{\text{no IRS}}$ and ${\rm SNR}_b^{\text{mrt,IRS-PCA}}$ reveal that the occurrence of the IRS-PCA leads to the decrease in the SNR of Bob, which is because $\bm{w}_{\rm mrt}$ in this case does not accurately match with Bob's channel $\bm{h}_b$.
Finally, by using the proposed cooperative channel estimation and beamforming scheme, Alice is able to reduce the signal leakage to the IRS, which causes two results: 1) compared with the naive MRT beamforming scheme, the SNR of Bob can be improved and approaches to ${\rm SNR}_b^{\text{no IRS}}$ and 2) the signal reflected by the IRS becomes weak and Eve benefits less from the IRS.

In Fig. \ref{STSNRvsJ}, we illustrate the SNRs of Bob and Eve versus the number of cooperative nodes $J$, where we set $M = 128$.
It can be seen from Fig. \ref{STSNRvsJ} that increasing $J$ improves ${\rm SNR}_{b}^{\text{zf,IRS-PCA}}$ and greatly reduces ${\rm SNR}_{e}^{\text{zf,IRS-PCA}}$.
This is because the estimation error of $\bar{\bm{h}}_{\rm I}$ gets small when $J$ increases as indicated by Theorem \ref{PropAccuracyOfChannelEstimation}.
If $J$ is sufficiently large, the proposed beamforming scheme restricts ${\rm SNR}_{e}^{\text{zf,IRS-PCA}}$ to be even smaller than ${\rm SNR}_{e}^{\text{IRS}}$ meaning that it is better for Eve not to perform IRS-PCA.

The amplitude of the reflecting coefficient at the IRS, $r_p$, plays an important role in the secrecy performance of the considered system. We illustrate the effects of $r_p$ on the SNRs of Bob and Eve in Fig. \ref{STSNRvsrp}, where we set $M = 128$.
In general, if Alice does not adopt any countermeasure on IRS-PCA, then $r_p = 1$ is optimal for Eve in term of maximizing its wiretapping SNR, as shown by the curve of ${\rm SNR}_e^{\text{mrt,IRS-PCA}}$.
However, if Alice utilizes the propose cooperative channel estimation and beamforming scheme, the case becomes different.
One one hand, if $r_p$ is large,  Alice can accurately estimate $\bar{\bm{h}}_{\rm I}$ and thus reduces Eve's SNR. On the other hand, if $r_p$ is too small, then the signal leakage from Alice to the IRS becomes small, and Eve benefits little from its IRS-PCA.
As a result, if Eve aims at maximizing its SNR, there usually exists an optimal value of $r_p$ within $(0,1)$, which can be seen from the curve of ${\rm SNR}_e^{\text{zf,IRS-PCA}}$.

In Fig. \ref{STSNRvsNtidle}, we assume that Eve selects the optimal value of $r_p$ to perform IRS-PCA (which is obtained by one-dimensional search with step size $0.025$ in our simulation), and the resulting SNRs of Bob and Eve are plotted versus the number of reflecting elements of Eve's IRS.
In Fig. \ref{STSNRvsNtidle}, by comparing the curve of ${\rm SNR}_e^{\text{mrt,IRS-PCA}}$ with $r_p = 1$ and the curve of ${\rm SNR}_e^{\text{zf,IRS-PCA}}$ with $r_p = r_{p,opt}$, we can conclude that the proposed secure transmission scheme can indeed effectively reduce the wiretapping SNR even if Eve is able to use the optimal reflecting amplitude. However, Fig. \ref{STSNRvsNtidle} also reveals  that if Eve is able to increase the number of reflecting elements of its IRS, then it can greatly enhance the wiretapping capability.
In practice, it is possible that Eve utilizes massive reflecting elements to attack the legitimate system, which poses severe threat to the security of the legitimate communication system. One possible solution is to increasing the number of cooperative nodes, which has been shown to be useful in Fig. \ref{STSNRvsJ}. Another possible solution is to design artificial noise transmission scheme to increase the noise power level of Eve, which is, however, out the scope of this paper and left for future work.

\section{Discussion \& Conclusions}
In this paper, we have proposed a new PCA scheme for Eve, namely IRS-PCA, wherein Eve uses an IRS to reflect the pilot sequence of Bob during the RPT phase. We extensively reviewed existing countermeasures on PCA in literature and showed that the proposed IRS-PCA scheme disables many existing methods. To combat with IRS-PCA, we proposed a sequential detection scheme, i.e., the GCUSUM scheme, for Alice to detect whether IRS-PCA has occurred. The ARL2FA, WADD, and WAWTG of the proposed GCUSUM scheme have been analyzed. Numerical experiments revealed that the proposed GCUSUM scheme is better than the benchmark ED scheme when the antenna number of Alice is not large enough or when a long ARL2FA is desired. To enable secure transmission under IRS-PCA, a cooperative channel estimation scheme has been proposed for Alice to estimate its channel to the IRS, which can be used to construct ZF beamforming vector to reduce the signal leakage. Numerical results were presented to show the performance of the proposed secure transmission scheme.
\section*{Acknowledgment}
K.-W Huang would like to thank the China Scholarship Council for the financial support.

\appendix
\subsection{The proof of Theorem 1}
\label{App:DetectionPerformance}
To prove Theorem 1,  the following two lemmas are required.
\begin{lemma}[Theorem 2, \cite{G.Lorden1971}, Proposition 2, \cite{Q.YaoSALinearNodel1993}]
{\em Let $T$ be a stopping time with respect to $\{\bm{y}_1,\bm{y}_2,\cdots\}$ such that $\mathbb{P}\{T<\infty\} \leq \frac{1}{\gamma}$ for some  $\gamma\in (1,\infty)$. For $\forall k \geq 1$, let $T^{(k)}$ denote the stopping time obtained by applying $T$ to $\{\bm{y}_k,\bm{y}_{k+1},\cdots\}$, and define $T^* = \min_{k\geq 1} \{ T^{(k)} + k - 1\}$. Then $T^*$ is a stopping time with $\mathbb{E}_{\infty}(T^*)\geq \gamma$ and $\mathcal{D}(T^*)\leq \sup_{k\geq 1}\mathbb{E}_k\left(T^{(k)}\right)$.}
\end{lemma}
\begin{lemma}
\label{ProofARL2FARequired2}
{\em Let $T_x = \inf\{n:n\geq 1,\frac{x}{1 + x}\sum_{k=1}^n\left(\|\bm{y}_k\|^2 - M\ln(1 + x)\right)\geq \eta\}$, where $x>0$, be a stopping time respect to $\{\bm{y}_1,\bm{y}_2,\cdots\}$. Under the condition that $\bm{y}\sim\mathcal{CN}(\bm{0},\bm{I}_M)$, $T_x$ satisfies $\mathbb{P}\{T_x < \infty\} \leq \mathrm{e}^{-\eta}$. }
\end{lemma}
\begin{IEEEproof}
Define $\mathcal{F}_0$ as the hypothesis $\{\bm{y}_k\sim \mathcal{CN}(\bm{0},\bm{I}_M) , \forall k\geq 1\}$ and $\mathcal{F}_1$ as the hypothesis $\{ \bm{y}_k\sim \mathcal{CN}(\bm{0},(1 + x)\bm{I}_M), \forall k\geq 1\}$. $T_x$ is in fact the one-sided sequential probability ratio test that tests $\mathcal{F}_0$ against $\mathcal{F}_1$ \cite{H.V.Poor2009}, and this lemma directly follows from \cite[Proposition 4.10]{H.V.Poor2009}.
\end{IEEEproof}

Based on Lemma 1, we now consider the following stopping time on $\{\bm{y}_1,\bm{y}_2,\cdots\}$
\begin{align}
T = \inf \left\{ n: n\geq 1, \tilde{\Lambda}_{n}(\bar{\xi}) > \eta \right\},
\end{align}
where for some $\bar{\xi}>0$, $\tilde{\Lambda}_{n}(\bar{\xi})$ is given by
\begin{align}
\tilde{\Lambda}_{n}(\bar{\xi}) &=
\sup_{\theta \geq \bar{\xi}}  \frac{\theta}{1 + \theta}S_n - nM  \ln \left(1 + \theta\right) \nonumber \\
&=
\left\{\begin{aligned}
& nM \left(\bar{S}_{n} - \ln \bar{S}_{n} - 1\right), &&  \text{ if }\bar{S}_{n} - 1 \geq \bar{\xi}, \\
& nM \left(\frac{\bar{\xi} \bar{S}_{n} }{ 1 + \bar{\xi} }  -  \ln \left(1 + \bar{\xi}\right)\right), &&  \text{ if }  \bar{S}_{n} - 1 < \bar{\xi},
\end{aligned}\right. \nonumber
\end{align}
with $S_n = S_{1,n}$ and $\bar{S}_{n} = \bar{S}_{1,n}$. It is obvious that $T_{\rm G} = T^*$. Therefore, to prove $\mathbb{E}_{\infty}(T_G)\geq \gamma$, we only need to prove $\mathbb{P}_{\infty}\{T <\infty\} \leq \frac{1}{\gamma}$.
\subsubsection{The proof of $\mathbb{P}_{\infty}\{T<\infty\} \leq \frac{1}{\gamma}$}
First of all, we have that
\begin{align}
&\sup_{\theta \geq \bar{\xi}}  \frac{\theta}{1 + \theta}S_n - nM  \ln \left(1 + \theta\right) \geq \eta\nonumber \\
\Leftrightarrow&
S_n \geq \inf_{\theta \geq \bar{\xi}}  \frac{ 1 + \theta}{\theta} \eta + \frac{ 1 + \theta}{\theta}  nM  \ln \left(1 + \theta\right) \nonumber \\
\Leftrightarrow&
S_n \geq \inf_{t \geq \bar{\xi}^{\prime} } \frac{ \eta }{ t } - \frac{ nM  \ln \left(1 - t\right)}{t}
= n \inf_{t \geq \bar{\xi}^{\prime} } f_n(t),\label{DetectionDelayDerivative}
\end{align}
where $t = \frac{\theta}{1 + \theta}\in(0,1)$, $\bar{\xi}^{\prime} = \frac{\bar{\xi}}{1 + \bar{\xi}}$, and $f_n(t) \triangleq \frac{ (\eta/n) -  M \ln \left(1 - t\right)}{ t }$. Now, we calculate the $\inf$ in \eqref{DetectionDelayDerivative}.
The derivative of $f_n(t)$ is given by
$f_n^{\prime}(t) = \frac{M\left( \frac{t}{1 - t} +  \ln \left(1 - t\right) \right)- (\eta/n)}{t^2}
= \frac{Mg(t)- (\eta/n)}{t^2}$,
where $g(t) \triangleq  \frac{t}{1 - t} +  \ln \left(1 - t\right)$.
Due to the fact that for $\forall t\in(0,1)$,
$g^{\prime}(t) = \frac{1}{(1 - t)^2} - \frac{1}{1 - t} = \frac{t}{(1 - t)^2} \geq 0$,
we obtain $0 = g(0)\leq g(t) \uparrow \infty$ as $t\rightarrow 1$, which means that there exists $t_n\in(0,1)$ such that
\begin{align}
\label{ffuncproperty}
\left\{\begin{aligned}
Mg(t_n) &= \frac{\eta}{n};\\
f_n^{\prime}(t) &\leq 0, {\rm ~for ~}t\leq t_n; \\
f_n^{\prime}(t) &\geq 0,~ {\rm ~for ~}t \geq  t_n.
\end{aligned}\right.
\end{align}
As $g(t)$ is monotonically increasing, we conclude that $t_1 > t_2 >t_3 > \cdots $. Define $n^*$ as follow
\begin{align}
\bar{n}  &= \inf_{n \geq 1} \left\{ n :  t_n \leq  \bar{\xi}^{\prime} \right\} = \left\lceil \frac{\eta}{Mg\left(\bar{\xi}^{\prime}\right)} \right\rceil \leq  \frac{\eta}{Mg\left(\bar{\xi}^{\prime}\right)} + 1.
\end{align}
Based on \eqref{ffuncproperty}, we have that $\inf_{t\geq \bar{\xi}^{\prime}} f_n(t) = f_n(t_n) $ if $n<\bar{n}$, and $\inf_{t\geq \bar{\xi}^{\prime}} f_n(t) = f_n(\bar{\xi}^{\prime}) $ if $n\geq \bar{n}$.
For event $\{n^* \leq  T < \infty\}$, we have that
\begin{align}
&\quad \mathbb{P}_{\infty}\left\{ n^* \leq  T < \infty \right\} \nonumber \\
&= \mathbb{P}_{\infty}\left\{\forall n < n^*, S_n  \leq f_n(t_n) ;  \exists n\geq n^*, S_n > f_n\left(\bar{\xi}^{\prime}\right) \right\} \nonumber \\
&\leq \mathbb{P}_{\infty}\left\{ \exists n\geq n^*, S_n  > f_n\left(\bar{\xi}^{\prime}\right) \right\} \nonumber \\
&\leq \mathbb{P}_{\infty}\left\{ \exists n\geq 1, S_n  > f_n\left(\bar{\xi}^{\prime}\right) \right\}\nonumber \\
&= \mathbb{P}_{\infty}\left\{ T_{\bar{\xi}} < \infty \right\} \overset{(*)}{\leq} \mathrm{e}^{-\eta}.
\end{align}
where step $(*)$ is due to Lemma \ref{ProofARL2FARequired2}. For $n < n^*$, we have that
\begin{align}
\mathbb{P}_{\infty}\left\{ T = n \right\} &= \mathbb{P}_{\infty}\left\{\forall k < n, S_k \leq f_k(t_k) ;   S_n > f_n(t_n) \right\} \nonumber \\
&\leq \mathbb{P}_{\infty}\left\{ T_{t_n^{\prime}} < \infty \right\} \overset{(*)}{\leq}  \mathrm{e}^{-\eta}.
\end{align}
where $t_n^{\prime} = \frac{t_n}{1 - t_n}$ and step $(*)$ is due to Lemma \ref{ProofARL2FARequired2}. Note that $\bar{n} \leq  \frac{h}{m g(\bar{\xi}^{\prime})} + 1$, therefore
$P\left\{ T  < \bar{n} \right\} \leq \frac{\eta}{M g\left(\bar{\xi}^{\prime}\right)} \mathrm{e}^{-\eta}$.
Therefore,
\begin{align}
\mathbb{P}_{\infty}\left\{ T  < \infty \right\} & \leq  \left(   \frac{\eta}{M g \left(\bar{\xi}^{\prime}\right)} + 1
\right)\mathrm{e}^{-\eta} \nonumber \\
& \leq \left(   2\frac{\eta}{M (\bar{\xi}^{\prime})^2} + 1 \right)\mathrm{e}^{-\eta}.
\end{align}
By selecting $\bar{\xi} = \frac{1}{\sqrt{M}\ln(\gamma)}$ and $\eta = (1+o_\gamma(1))\ln \gamma $, we have $\mathbb{P}_{\infty}\left\{ T  < \infty \right\} \leq  \frac{1}{\gamma}$ for $\gamma\rightarrow \infty$.

\subsubsection{The proof of $\mathcal{D}(T_{\rm G}) = \mathcal{O} \left(  \frac{\ln \gamma}{M( \mu - \ln(1 + \mu) )} \right) $}
Based on Lemma 1, we only need to prove that
$\mathbb{E}_k \left( T^{(k)} \right) =  \mathcal{O} \left(  \frac{ \ln\gamma }{M(\mu - \ln(1 + \mu))} \right) $ for $\forall k\geq 1$.
In the following, for simplicity, we only prove this statement for the case with $k = 1$, i.e., $\mathbb{E}_1 \left( T^{(1)} \right) = \mathbb{E}_1 \left( T \right) \leq \frac{(1 + o_{\gamma}(1))\ln\gamma }{M(\mu - \ln(1 + \mu))}$. Using the same approach, this conclusion can be easily extended to the case with $k > 1$ due to the uniform convergence in Assumption \ref{AverageAttackPowerAssumption}.

Note that $\bar{\xi} \rightarrow 0$ as $\gamma\rightarrow \infty$, and thus $1 + \mu>1 + \bar{\xi}$ as $\gamma\rightarrow \infty$. Based on Kolmogorov's strong law of large numbers, we have
$ \bar{S}_n \xrightarrow[n\rightarrow\infty]{a.s.} 1 + \mu > 1 + \bar{\xi}$ under the condition that $\nu = 1$. Using the fact that $T\rightarrow\infty$ as $\eta\rightarrow \infty$, we have
\begin{align}
T = \inf \left\{ n: n\geq 1,  nM \Delta(\bar{S}_{n}) > \eta \right\},{\rm ~as~} \eta \rightarrow \infty,
\end{align}
where $\Delta(x) \triangleq x - \ln x - 1$.
Note that $\mathbb{P}_1\left\{1 + \frac{1}{2}\mu \leq \bar{S}_{n} \leq 1 + \frac{3}{2}\mu\right\}\xrightarrow{n\rightarrow \infty} 1$, and on event $\big\{1 + \frac{1}{2}\mu \leq \bar{S}_{n} \leq 1 + \frac{3}{2}\mu\big\}$, we have
\begin{align}
&\quad\  nM \Delta(\bar{S}_{n}) \nonumber \\
&= nM \Big\{\mu - \ln (1 + \mu) + \frac{\mu}{1 + \mu}\left(\bar{S}_{n} - (1 + \mu)\right)\nonumber \\
&\quad\quad\quad\quad\quad\quad\quad\quad\quad\quad\quad\quad\
+ \frac{1}{2 v_n^2}\left(\bar{S}_{n} - (1 + \mu)\right)^2  \Big\} \nonumber \\
&= nM \left\{\frac{\mu}{1 + \mu} \bar{S}_{n} - \ln (1 + \mu) + \frac{1}{2 v_n^2}\left(\bar{S}_{n} - (1 + \mu)\right)^2\right\} \nonumber \\
&> nM \left\{\frac{\mu}{1 + \mu} \bar{S}_{n} - \ln (1 + \mu) \right\}\nonumber \\
&= \sum_{k=1}^n \left(\frac{\mu}{1 + \mu}  \frac{\left\|\bm{y}_k\right\|^2}{\sigma_0^2}- M\ln (1 + \mu)\right),
\end{align}
where we expand $\Delta(x)$ around $x = 1 + \mu$ and $v_n$ is a random variable distributed  within $(1 + \frac{1}{2}\mu,1 + \frac{3}{2}\mu)$.
Therefore, as $\eta\rightarrow \infty$, we have that
\begin{align}
T < \tilde{T} &  \triangleq
\inf \Bigg\{ n: n\geq 1, \nonumber \\
& \sum_{k=1}^n \left(\frac{\mu}{1 + \mu}  \frac{\left\|\bm{y}_k\right\|^2}{\sigma_0^2}- M \ln (1 + \mu)\right)  > \eta \Bigg\},
\end{align}
Using Assumption \ref{AverageAttackPowerAssumption}, we have that $\frac{1}{n}\sum_{k=1}^n \left(\frac{\mu}{1 + \mu}  \frac{\mathbb{E}\left(\left\|\bm{y}_k\right\|^2\right)}{\sigma_0^2}- M \ln (1 + \mu)\right)\rightarrow M(\mu - \ln (1 + \mu) ) $ as $n\rightarrow \infty$. Based on \cite[Theorem 2]{Y.S.Chow}, we have that
$\lim_{\eta \rightarrow \infty} \frac{\mathbb{E}_1 (\tilde{T}) }{ \eta } =  \frac{1}{M \left(\mu - \ln (1 + \mu)\right)}$.
As $\eta = (1 + o_{\gamma}(1)) \ln(\gamma)$, we obtain that $\mathbb{E}_1 (T^{(1)}) =  \mathbb{E}_1 (T) = \mathcal{O} \left(  \frac{ \ln \gamma}{M \left(\mu - \ln (1 + \mu)\right)}\right) $ as $\gamma\rightarrow \infty$.

\subsection{The Proof of Corollary 2 and 3}
\label{Appendix:WAWTG}
First of all, we note that under the condition $\bm{\Phi}^{(d)} = \bm{\Phi}_{\nu+k}^{(d)}$ and $\bm{\Phi}^{(p)} = \bm{\Phi}_{\nu+k}^{(p)}$ for $\forall k \geq 0$,
$\left\{\Delta\mathcal{I}_{\nu+k}:k\geq 0\right\}$ is  sequence  of  i.i.d. random variables, and it can be easily checked that $\mathbb{E}(\Delta\mathcal{I}_{\nu+k})>0$ for $\forall k\geq 0$. $\mathcal{W}(T_G)$ can be upper bounded as follow,
\begin{align}
\mathcal{W}(T_G) &= \sup_{\nu \geq 1}\mathop{\rm esssup}\limits_{\bm{y}_1^{\nu-1}}
\mathbb{E}_{\nu}\left\{ {\rm WTG}_{T_G -1}|\bm{y}_1^{\nu-1} \right\}\nonumber \\
&\leq \sup_{\nu \geq 1}\mathop{\rm esssup}\limits_{\bm{y}_1^{\nu-1}}\mathbb{E}_{\nu}\left\{ {\rm WTG}_{T_G}| \bm{y}_1^{\nu-1} \right\} \nonumber \\
&\leq  \mathbb{E}_{\nu}\left\{ {\rm WTG}_{T_G} | \bm{y}_1^{\nu-1} = \bm{0} \right\}\nonumber \\
&\leq  \mathbb{E}_{\nu}\left\{ \sum_{k = \nu}^{T^{(\nu)} +\nu - 1} \Delta\mathcal{I}_{k} \right\}
= \mathbb{E}_{1}\left\{ \sum_{k = 1}^{T} \Delta\mathcal{I}_{k} \right\} \nonumber \\
&= \mathbb{E}_{1}(T)\mathbb{E}(\Delta\mathcal{I}_{1}),
\label{ProofC23}
\end{align}
where the last step is due to Wald's identity \cite[Corollary 2.3.1]{A.G.Tartakovsky2013}.
Combining \eqref{ProofC23} with Corollary 1 leads to Corollary 2 and 3.
\subsection{The Proof of Theorem \ref{PropAccuracyOfChannelEstimation}}
\label{Appendix:EstimationAccuracy}
Based on \eqref{EstimateHelperChannel}, we have
$\bm{T} =\left[\frac{1}{\sigma_1}\bm{t}_1,\frac{1}{\sigma_2}\bm{t}_2,\cdots,\frac{1}{\sigma_J}\bm{t}_J\right] = \left[\hat{\bm{t}}_1,\hat{\bm{t}}_2,\cdots,\hat{\bm{t}}_J\right]$ and
$\bm{T}\bm{T}^H  = \hat{\bm{t}}_1\hat{\bm{t}}_1^H + \hat{\bm{t}}_2\hat{\bm{t}}_2^H + \cdots + \hat{\bm{t}}_J\hat{\bm{t}}_J^H$.
Any eigenvector of $\bm{T}\bm{T}^H$ that corresponds to a non-zero eigenvalue, e.g., denoted by $\varrho$, should be in the form of $ \bm{T}\bm{\kappa}$ for $\bm{\kappa}\in\mathcal{C}^{J\times 1}$ and $\bm{\kappa} \neq \bm{0}$, then
\begin{align}
&\bm{T}\bm{T}^H\bm{T}\bm{\kappa} = \varrho \bm{T}\bm{\kappa} \nonumber \\
~\Rightarrow~&\bm{T}^H\bm{T}\bm{T}^H\bm{T}\bm{\kappa} = \varrho \bm{T}^H\bm{T}\bm{\kappa}\nonumber \\
~\overset{\rm a.s.}{\Rightarrow}~ &
\bm{T}^H\bm{T}\bm{\kappa} = \varrho \bm{\kappa},
\end{align}
where the last step is because $\mathbb{P}\{{\rm rank}(\bm{T}^H\bm{T}) = J\} = 1$ in our case. Then $\bm{\kappa}$ is a eigenvector of $\bm{T}^H\bm{T}$ with the corresponding eigenvalue being $\varrho$. Note that as $M\rightarrow \infty$, it can be easily verified that $\frac{1}{M}\bm{T}^H\bm{T} \rightarrow \bm{I}_J + \bm{a}_{\rm CN}^*\bm{a}_{\rm CN}^T$.
Therefore, $\bm{\kappa}\rightarrow\bm{a}_{\rm CN}^*$ and $\bm{\lambda}_{\max}(\bm{T}\bm{T}^H)\rightarrow
\bm{T}\bm{a}_{\rm CN}^*/\|\bm{T}\bm{a}_{\rm CN}^*\|$ with $\varrho \rightarrow 1 + \|\bm{a}_{\rm CN}\|^2$.
Furthermore, we have
\begin{align}
\label{Theorem:3ProofFinalStep}
\bm{T}\bm{a}_{\rm CN}^* &= \sum_{j=1}^J \left(\tilde{\bm{f}}_j + \frac{a_j}{\sigma_j} \bm{h}_{\rm I}\right)\frac{a_j^*}{\sigma_j}\nonumber \\
&=\left(\sum_{j=1}^J\frac{|a_j|^2}{\sigma_j^2}\right)\bm{h}_{\rm I} + \sum_{j=1}^J \frac{a_j^*}{\sigma_j} \tilde{\bm{f}}_j
\end{align}
where $\tilde{\bm{f}}_j = \frac{1}{\sigma_j}\left(\bm{f}_j + \tilde{\bm{z}}_j\right)\sim\mathcal{CN}(\bm{0},\bm{I}_M)$. Dividing the both sides of \eqref{Theorem:3ProofFinalStep} by $\|\bm{a}_{\rm CN}\|^2$ leads to the result in Theorem 3.

\subsection{Proof of Theorem \ref{Therorem:SNREAfterZF}}
\label{Appendix:ProofSNREAfterZF}
Due to the fact that $\bm{h}_e$ is independent of $\bm{y}$ and $\hat{\bar{\bm{h}}}_{\rm I}$, we have $\mathbb{E}\{\mathrm{SNR}_e^{\rm \text{zf,IRS-PCA}}\} = \frac{a_e^2}{\sigma_e^2} + \frac{|\tilde{a}_{{\rm I}}|^2}{\sigma_e^2}
\mathbb{E} \left\{\frac{\mathfrak{N}}
{\mathfrak{D}}\right\}$, where
where
$\mathfrak{N}\triangleq \frac{1}{M}\left|\bm{h}_{{\rm I}}^H
\left(\bm{I}_M - \hat{\bar{\bm{h}}}_{\rm I}\hat{\bar{\bm{h}}}_{\rm I}^H \right)
\bm{y}\right|^2$ and
$\mathfrak{D}\triangleq \frac{1}{M}\left\|\left(\bm{I}_M - \hat{\bar{\bm{h}}}_{\rm I}\hat{\bar{\bm{h}}}_{\rm I}^H \right)\bm{y}\right\|^2$.

Denote $\tilde{\bm{h}}_{b} = \bm{h}_b + \bm{z}$. For $\mathfrak{N}$, we further have that
\begin{align}
\mathfrak{N}& =\frac{|\hat{a}_{{\rm I} } |^2}{M}\left|
\bm{h}_{{\rm I}}^H \left(\bm{I}_M - \hat{\bar{\bm{h}}}_{\rm I}\hat{\bar{\bm{h}}}_{\rm I}^H \right)\bm{h}_{{\rm I} }\right|^2\nonumber \\
&\quad  +
\frac{1}{M}\left|\bm{h}_{{\rm I}}^H
\left(\bm{I}_M - \hat{\bar{\bm{h}}}_{\rm I}\hat{\bar{\bm{h}}}_{\rm I}^H \right)\tilde{\bm{h}}_{b}\right|^2 \nonumber \\
& \quad  + \frac{2}{M}\Re\left\{\hat{a}_{{\rm I} } \tilde{\bm{h}}_{b}^H
\left(\bm{I}_M - \hat{\bar{\bm{h}}}_{\rm I}\hat{\bar{\bm{h}}}_{\rm I}^H \right)
\bm{h}_{\rm I}
\right\}\nonumber \\
&\quad\times
\bm{h}_{\rm I}^H
\left(\bm{I}_M - \hat{\bar{\bm{h}}}_{\rm I}\hat{\bar{\bm{h}}}_{\rm I}^H \right)
\bm{h}_{\rm I}.\nonumber
\end{align}
Similarly, for $\mathfrak{D}$, we further have that
\begin{align}
\mathfrak{D}&
=\frac{1}{M} \left(\tilde{\bm{h}}_{b} + \hat{a}_{{\rm I} } \bm{h}_{{\rm I} } \right)^H
\left(\bm{I}_M - \hat{\bar{\bm{h}}}_{\rm I}\hat{\bar{\bm{h}}}_{\rm I}^H \right)
\left(\tilde{\bm{h}}_{b} + \hat{a}_{{\rm I} } \bm{h}_{{\rm I} } \right) \nonumber \\
&= \frac{1}{M} \tilde{\bm{h}}_{b}^H\left(\bm{I}_M - \hat{\bar{\bm{h}}}_{\rm I}\hat{\bar{\bm{h}}}_{\rm I}^H \right)\tilde{\bm{h}}_{b}
+
\frac{|\hat{a}_{{\rm I}}|^2}{M} \bm{h}_{{\rm I} }^H\left(\bm{I}_M - \hat{\bar{\bm{h}}}_{\rm I}\hat{\bar{\bm{h}}}_{\rm I}^H \right)\bm{h}_{{\rm I} }\nonumber \\
&\quad + \frac{2}{M}\Re \left\{\hat{a}_{{\rm I}}
\tilde{\bm{h}}_{b}^H\left(\bm{I}_M - \hat{\bar{\bm{h}}}_{\rm I}\hat{\bar{\bm{h}}}_{\rm I}^H \right)\bm{h}_{{\rm I} }\right\}.
\nonumber
\end{align}
Using the fact that $\hat{\bar{\bm{h}}}_{\rm I}\xrightarrow{M\rightarrow\infty} \frac{\bm{h}_{\rm I}+\bm{e}}{\|\bm{h}_{\rm I} + \bm{e}\|}$, we can obtain  that
$\frac{1}{M}\bm{h}_{\rm I}^H (\bm{I}_M - \hat{\bar{\bm{h}}}_{\rm I}\hat{\bar{\bm{h}}}_{\rm I}^H  )\bm{h}_{\rm I}
\rightarrow  \frac{1}{1 + \|\bm{a}_{\rm CN}\|^2}$,
$\frac{1}{M}  | \tilde{\bm{h}}_{b}^H (\bm{I}_M - \hat{\bar{\bm{h}}}_{\rm I}\hat{\bar{\bm{h}}}_{\rm I}^H  )\bm{h}_{{\rm I} } |^2\rightarrow \mathcal{O}(1)$,
$\frac{1}{M} \tilde{\bm{h}}_{b}^H (\bm{I}_M - \hat{\bar{\bm{h}}}_{\rm I}\hat{\bar{\bm{h}}}_{\rm I}^H  )\bm{h}_{{\rm I} } \rightarrow \mathcal{O}(1/\sqrt{M})$, and
$\frac{1}{M}\tilde{\bm{h}}_{b}^H (\bm{I}_M - \hat{\bar{\bm{h}}}_{\rm I}\hat{\bar{\bm{h}}}_{\rm I}^H )\tilde{\bm{h}}_{b}\rightarrow \sigma_0^2$.
Combining these results together, we obtain that $\mathfrak{N} \rightarrow
|\hat{a}_{{\rm I}}|^2 \left(\frac{1}{1 + \|\bm{a}_{\rm CN}\|^2}\right)^2 M$ and $\mathfrak{D}\rightarrow\sigma_0^2 + \frac{|\hat{a}_{{\rm I}}|^2}{1 + \|\bm{a}_{\rm CN}\|^2}$, which proves Theorem 4.

\end{document}